\documentclass[reprint,superscriptaddress,amsmath,longbibliography,amssymb,prl]{revtex4-2}

\usepackage{graphicx}
\usepackage{dcolumn}
\usepackage{bm}
\usepackage{multirow}
\usepackage{array}

\usepackage[usenames,dvipsnames]{xcolor}
\definecolor{nblue}{rgb}{0.0, 0.0, 1.0}
\definecolor{magenta}{rgb}{0.79, 0.08, 0.48}
\usepackage[colorlinks,linkcolor=nblue,urlcolor=magenta,citecolor=magenta,plainpages=false,pdfpagelabels,breaklinks]{hyperref}

\newcommand{\beq}{\begin{equation}}
\newcommand{\eeq}{\end{equation}}
\newcommand{\bea}{\begin{eqnarray}}
\newcommand{\eea}{\end{eqnarray}}

\begin{document}

\title{de Haas-van Alphen effect and the first-principles study of the possible topological stannide Cu$_3$Sn}

\author{Chengxu Liu}
\thanks{These authors contribute equally to this work.}
\affiliation{Key Laboratory of Quantum Precision Measurement of Zhejiang Province, Department of Applied Physics, Zhejiang University of Technology, Hangzhou 310023, China}

\author{Bin Li}
\thanks{These authors contribute equally to this work.}
\affiliation{New Energy Technology Engineering Laboratory of Jiangsu Province and School of Science, Nanjing University of Posts and Telecommunications, Nanjing 210023, China}

\author{Yongheng Ge}
\thanks{These authors contribute equally to this work.}
\affiliation{Key Laboratory of Low-Dimensional Quantum Structures and Quantum Control of Ministry of Education, Department of Physics and Synergetic Innovation Center for Quantum Effects and Applications, Hunan Normal University, Changsha 410081, China}

\author{Wen-He Jiao}
\email[]{whjiao@zjut.edu.cn}
\affiliation{Key Laboratory of Quantum Precision Measurement of Zhejiang Province, Department of Applied Physics, Zhejiang University of Technology, Hangzhou 310023, China}

\author{Chuanying Xi}
\affiliation{Anhui Province Key Laboratory of Condensed Matter Physics at Extreme Conditions, High Magnetic Field Laboratory,
Chinese Academy of Sciences, Hefei 230031, China}

\author{Yi Liu}
\affiliation{Key Laboratory of Quantum Precision Measurement of Zhejiang Province, Department of Applied Physics, Zhejiang University of Technology, Hangzhou 310023, China}

\author{Chunqiang Xu}
\affiliation{School of Physics and Key Laboratory of MEMS of the Ministry of Education, Southeast University, Nanjing 211189, China}
\affiliation{Department of Physics and Astronomy, Michigan State University, East Lansing, Michigan 48824-2320, USA}

\author{Qi Lu}
\affiliation{Key Laboratory of Artificial Structures and Quantum Control (Ministry of Education),
Shenyang National Laboratory for Materials Science, School of Physics and Astronomy,
Shanghai Jiao Tong University, Shanghai 200240, China}

\author{Yunlong Li}
\affiliation{Key Laboratory of Artificial Structures and Quantum Control (Ministry of Education),
Shenyang National Laboratory for Materials Science, School of Physics and Astronomy,
Shanghai Jiao Tong University, Shanghai 200240, China}

\author{Hang-Qiang Qiu}
\affiliation{Department of Applied Physics, Zhejiang University of Science and Technology, Hangzhou 310023, China}

\author{Qin-Qing Zhu}
\affiliation{Key Laboratory for Quantum Materials of Zhejiang Province, School of Science,
Westlake University, Hangzhou 310024, China}
\affiliation{Institute of Natural Sciences, Westlake Institute for Advanced Study, Hangzhou 310024, China}

\author{Zhi Ren}
\affiliation{Key Laboratory for Quantum Materials of Zhejiang Province, School of Science,
Westlake University, Hangzhou 310024, China}
\affiliation{Institute of Natural Sciences, Westlake Institute for Advanced Study, Hangzhou 310024, China}

\author{Ziming Zhu}
\affiliation{Key Laboratory of Low-Dimensional Quantum Structures and Quantum Control of Ministry of Education, Department of Physics and Synergetic Innovation Center for Quantum Effects and Applications, Hunan Normal University, Changsha 410081, China}

\author{Dong Qian}
\affiliation{Key Laboratory of Artificial Structures and Quantum Control (Ministry of Education),
Shenyang National Laboratory for Materials Science, School of Physics and Astronomy,
Shanghai Jiao Tong University, Shanghai 200240, China}
\affiliation{Tsung-Dao Lee Institute, Shanghai Jiao Tong University, Shanghai 200240, China}

\author{Xianglin Ke}
\affiliation{Department of Physics and Astronomy, Michigan State University, East Lansing, Michigan 48824-2320, USA}

\author{Xiaofeng Xu}
\email[]{xuxiaofeng@zjut.edu.cn}
\affiliation{Key Laboratory of Quantum Precision Measurement of Zhejiang Province, Department of Applied Physics, Zhejiang University of Technology, Hangzhou 310023, China}

\date{\today}

\begin{abstract}
The quest for quantum materials with diverse symmetry-protected topological states has been the focus of recent research interest, primarily due to their fascinating physical properties and the potential technological utility. In this work, we report on the magnetotransport, de Haas-van Alphen (dHvA) oscillations, and the first-principles calculations of the stannide Cu$_3$Sn that is isostructural with the recently reported topological semimetal Ag$_3$Sn. The magnetoresistance was found to vary quasi-linearly in field. Clear dHvA oscillations were observed under a field as low as 1 Tesla at 2 K, with three major oscillation frequencies $F_{\alpha}$=8.74 T, $F_{\beta}$=150.19 T and $F_{\gamma}$=229.66 T and extremely small effective masses. The analysis of dHvA quantum oscillations revealed a possible nonzero Berry phase, suggestive of the nontrivial band topology. The corroborating evidence for the nontrivial electronic topology also comes from the first-principles calculations which yield a nonzero $\mathbb{Z}_2$ topological index. These results collectively suggest that Cu$_3$Sn, in analogy to its homologue Ag$_3$Sn, may be another intermetallic stannide hosting topological Dirac fermions.
\end{abstract}

\maketitle

\section{Introduction}
In recent years, there has been mounting research interest in topological semimetals, such as Dirac semimetals, Weyl semimetals or nodel-line semimetals, ever since the materials classification based on their symmetry and topology was extended from insulators to metals or semimetals \cite{armitage-review,ding-review}. The \textit{relativistic} fermions in these materials manifest themselves in a plethora of exotic physical phenomena, such as unsaturated linear magnetoresistance (MR) \cite{Ong}, the chiral anomaly effect \cite{Parame,Chen,PHE-Kumar}, nontrivial quantum oscillations \cite{He-prl,CXH-prl} and the anomalous Hall effect \cite{Liu-np}, etc., thereby opening up the avenue for vast material functionalities and applications in future devices and technologies. Among these topological semimetals, the class of binary stannides has attracted growing attention owing to their intriguing physical properties. For example, inspired by the observation of ultrahigh MR in the binary PtSn$_4$ \cite{PtSn4_MR}, the angle-resolved photoemission spectroscopy (ARPES) studies subsequently revealed a novel type of topological phase, namely, the surface-derived Dirac nodes that formed open arcs (rather than a closed loop) in reciprocal space~\cite{PtSn4_ARPES}. More interestingly, its sister compound PdSn$_4$ was reported to possess topological carriers that were significantly enhanced in the effective mass, possibly due to the the vanishing density of states near the Fermi level and the resultant weaker screening of the Coulomb interaction~\cite{PtSn4}. Remarkably, in some extreme settings, the topological carriers in some stannides may undergo superconducting transitions at low temperatures, thereby raising the possibility to realize the highly sought topological superconductivity in the family of stannides~\cite{AuSn4,CaSn3}.

In this broad context, the topological nodal-line semimetal AuSn$_4$ was recently reported to host peculiar two dimensional superconductivity with a transition temperature $T_c$ $\sim$ 2.40 K arising from its Dirac surface state, leading to the speculation that the topological superconductivity may exist on its surface~\cite{AuSn4}. In parallel, another stannide Ag$_3$Sn was also found to exhibit a nontrivial Berry phase in the Shubnikov-de Haas (SdH) oscillations and display a strange angle-dependent magnetoresistance, placing this binary alloy in the shrine of topological materials~\cite{ZN-prb}. A natural question arises as to the topological property of Cu-Sn alloys, since the elements of Au, Ag and Cu reside in the same group in the periodic table. In this work, we address this question in the affirmative.

Here, we report the synthesis of Cu-Sn alloy Cu$_3$Sn by the solid-state reaction method. Cu$_3$Sn grown in this way is found to crystallize in the same space group as the previously reported Ag$_3$Sn~\cite{ZN-prb}. Additionally, quantum dHvA oscillations suggest the nontrivial Berry phase of its underlying carriers; the nontrivial $\mathbb{Z}_2$ index and the Dirac-like surface spectrum revealed by the first-principles calculations are also consistent with the topologically nontrivial states in this material. Taken together, our results suggest that, similar to its analogues AuSn$_4$ and Ag$_3$Sn, the $Pmmn$ phase of Cu$_3$Sn may be another stannide that hosts topological Dirac fermions and therefore merits the in-depth investigations in the future.

\section{Methods}

High-quality Cu$_3$Sn single crystals were synthesized by the self-flux method. Starting materials Copper (Cu, 99.99 \%) powder and Tin (Sn, 99.99 \%) shots were mixed together in a molar ratio of Cu : Sn = 6 : 5 in a glove box filled with highly pure argon gas (O$_2$ and H$_2$O $<$ 0.1 ppm). The mixture was then sealed in an evacuated quartz tube. The quartz tube was heated to 1173 K and kept at this temperature for 10 hours to ensure that the mixture was melt thoroughly. The tube was subsequently cooled to 673 K at a rate of 2 K/h, followed by a furnace-cooling to room temperature. The shiny platelet-shaped Cu$_3$Sn crystals were finally obtained [see the photographic image shown in Fig.~\ref{fig1}]. The X-ray diffraction (XRD) data acquisition was performed at room temperature with a monochromatic Cu K$_{\alpha 1}$ radiation using a PANalytical x-ray diffractometer (Model EMPYREAN) radiation by a conventional $\theta$-2$\theta$ scan for a crystal mounted on the sample holder. Energy-dispersive x-ray spectroscopy (EDS) was carried out by the Hitachi S-3400 instrument to get the chemical composition of the as-grown crystals. The EDS analysis was performed on the fresh surface of the selected crystals, which gives the average composition of Cu$_{3.00}$Sn$_{0.97}$, very close to the stoichiometric Cu$_3$Sn. The detailed results are tabulated in Table S1 of the Supplemental Material (SM). The crystal structure was plotted with the software VESTA \cite{VESTA}. The Laue diffraction was carried out to confirm the crystal structure and the orientations of the crystals. The typical EDS spectrum for the crystal was shown in Fig.~\ref{fig1}(d). The (magneto-)transport measurements were performed in the Physical Property Measurement System (PPMS-9, Quantum Design) with the standard four-probe method. Magnetic susceptibility were measured on a commercial Quantum Design magnetic property measurement system (MPMS-7).

The first-principles calculations including spin-orbit coupling (SOC) were calculated by density functional theory implemented in the WIEN2K code~\cite{Wien2k}. The Perdew-Burke-Ernzerhof generalized gradient approximation (PBE-GGA)~\cite{GGA} was applied to the exchange-correlation potential calculation. The atomic positions were optimized by minimizing the forces and stress. The muffin tin radii were chosen to be 2.34 a.u.\ for Cu and 2.44 a.u.\ for Sn. The plane-wave cutoff was defined by $RK_{max}=7.0$, where $R$ is the minimum linearized augmented plane wave (LAPW) sphere radius and $K_{max}$ is the maximum plane-wave vector cutoff. We used Cu $s$ and $d$, Sn $s$ and $p$ orbitals to construct the Wannier functions for the calculation of surface state spectrum and topological indices. We performed the first-principles tight-binding model Hamilton, where the tight-binding model matrix elements were calculated by projecting onto the Wannier orbitals as implemented in WannierTools~\cite{wannier1,wannier2}.

\section{Results and Discussion}

\begin{figure*}
\begin{center}
\includegraphics[width=0.9\textwidth]{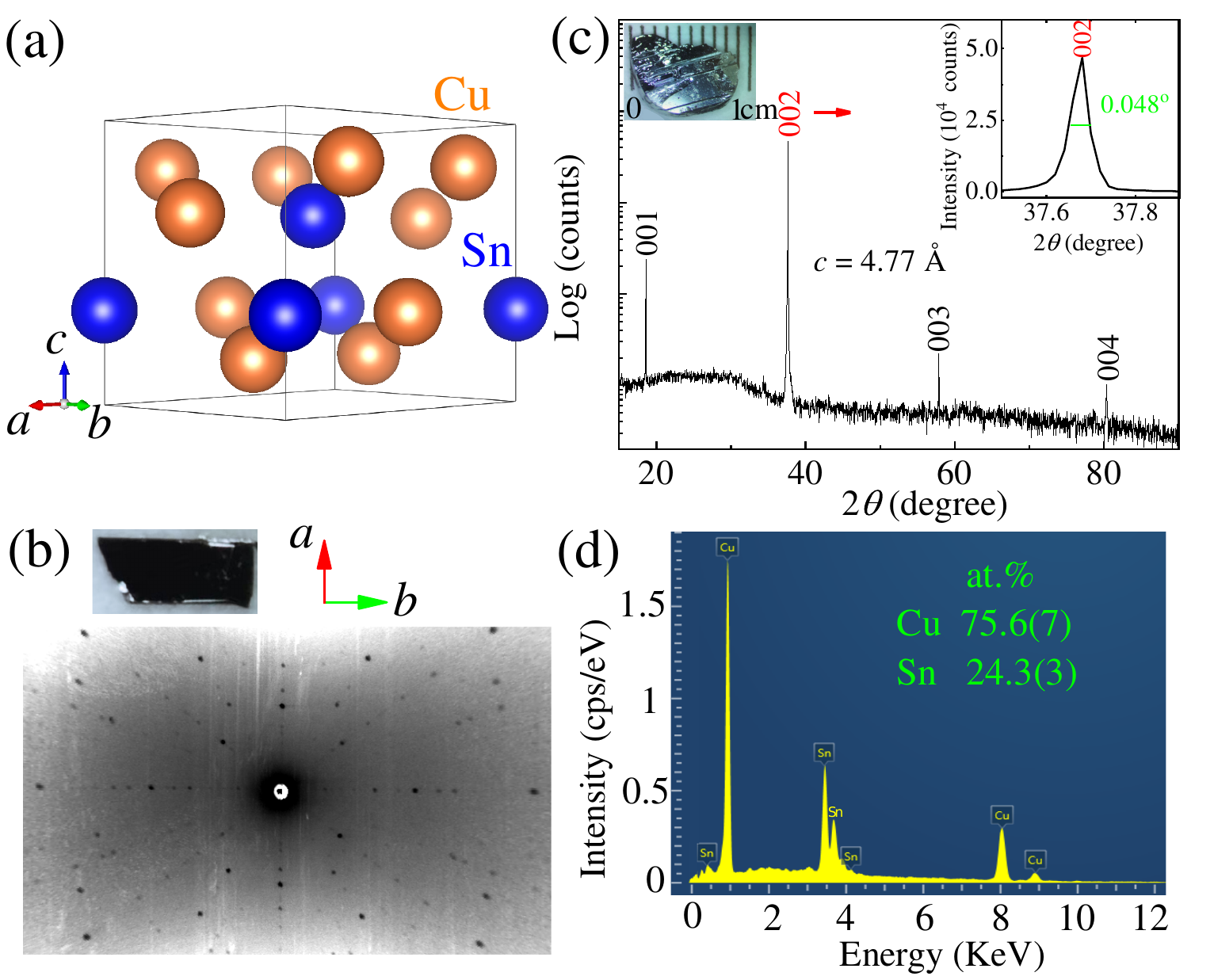}
\caption{\label{fig1}(Color online) Crystallographic structure and sample characterizations of Cu$_3$Sn. (a) The crystallographic structure of Cu$_3$Sn with an orthorhombic unit cell plotted using VESTA software. (b) Laue picture along the [001] axis (lower panel) and the as-grown crystal (upper panel) on which the Laue picture was taken. (c) Single-crystal x-ray diffraction pattern at room temperature. The right inset enlarges the second reflection in the x-ray diffraction pattern. The left inset is a photograph of the as-grown Cu$_3$Sn crystals. (d) A typical energy-dispersive x-ray spectrum with electron beams focused on the selected area (marked in the inset).}
\end{center}
\end{figure*}

\begin{figure*}
\begin{center}
\includegraphics[width=0.98\textwidth]{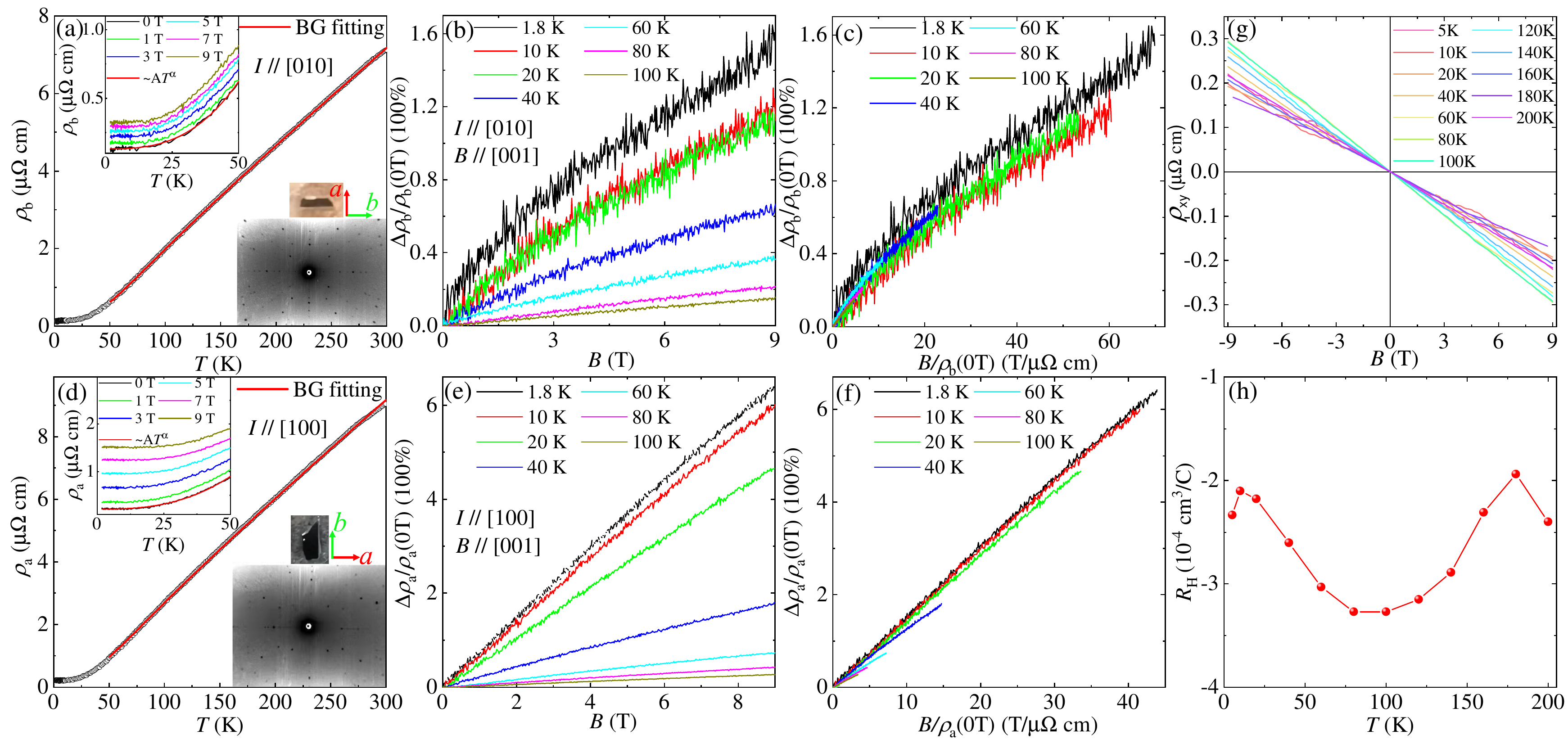}
\caption{\label{rt} (Color online) Temperature dependence of the electrical resistivity $\rho_{b}$ (a) and $\rho_{a}$ (d). The solid red line corresponds to fits to the BG equation. The right insets display the crystals used in measuring $\rho_{b}$ and $\rho_{a}$, and the Laue diffraction pictures used to find the orientation of the crystals before the measurements. The left insets magnify the low-$T$ dependence of $\rho_{b}$ and $\rho_{a}$ with the magnetic field parallel to the $c$-axis up to 9 T below 50 K. In zero field, $\rho_{b}$ can be fitted using a power-law dependence $T^\alpha$ ($\alpha$=2.85). For $\rho_a$, $\alpha$=2.68. The transverse magnetoresistance for $\rho_{b}$ (b) and $\rho_{a}$ (e) at selected temperatures. The orange solid lines are the fits (see main text). The Kohler's scaling for $\rho_{b}$ (c) and $\rho_{a}$ (f). (e) The Hall resistivity at different temperatures below 200 K. (f) The temperature dependence of Hall coefficients extracted from the Hall resistivity by linear fits.}
\end{center}
\end{figure*}

As reported in the literature, Cu$_3$Sn can crystallize in several distinct structures, including the space groups of $Pmmn$ \cite{Cu3Sn-59}, $F$\={4}3$m$ \cite{Cu3Sn-216}, $Pmm$2 \cite{Cu3Sn-25}, and $P$6$_3$/$mmc$ \cite{Cu3Sn-194}. The XRD pattern of the crystals with the plate-like facet lying on the sample holder at 298 K is shown in Fig.~\ref{fig1}(c). A set of diffraction peaks from (00$l$) can be observed. The interplanar spacing is calculated to be 4.78 {\AA}, very close to the reported lattice parameter of $c$ = 4.74 {\AA} for the orthorhombic $Pmmn$ phase \cite{Cu3Sn-59}. The full width at half-maximum of the diffraction is only 0.049$^\circ$, e.g., for the (002) peak, indicating high quality of our crystals. The Laue diffraction photograph along this axis, as shown in the lower panel of Fig.~\ref{fig1}(b), further indicates the single crystalline nature of the crystals. Meanwhile, we simulated the Laue diffraction patterns for all possible space groups quoted above in which Cu$_3$Sn can crystallize. As summarized in the SM, the simulated pattern for the $Pmmn$ phase shows the best consistency with the experimental one on the same scale, thereby corroborating the structural type of our crystals to be $Pmmn$, isostructural to the intermetallic stannide (IMS) Ag$_3$Sn \cite{ZN-prb}. As noted, on the shiny plate-like surface of the crystals, we can observe some striped lines [the left inset of Fig.~\ref{fig1}(c)] and striped boundaries [the upper panel of Fig.~\ref{fig1}(b)]. Our Laue diffraction measurements indicate those striped lines or boundaries are actually aligned along the crystallographic $b$ axis, which suggests the chemical bonding is much stronger along this direction. The crystallographic structure of Cu$_3$Sn is displayed in Fig.~\ref{fig1}(a), from which we see each unit cell comprised of six Cu atoms and two Sn atoms.

\begin{figure*}
\includegraphics[width=0.98\textwidth]{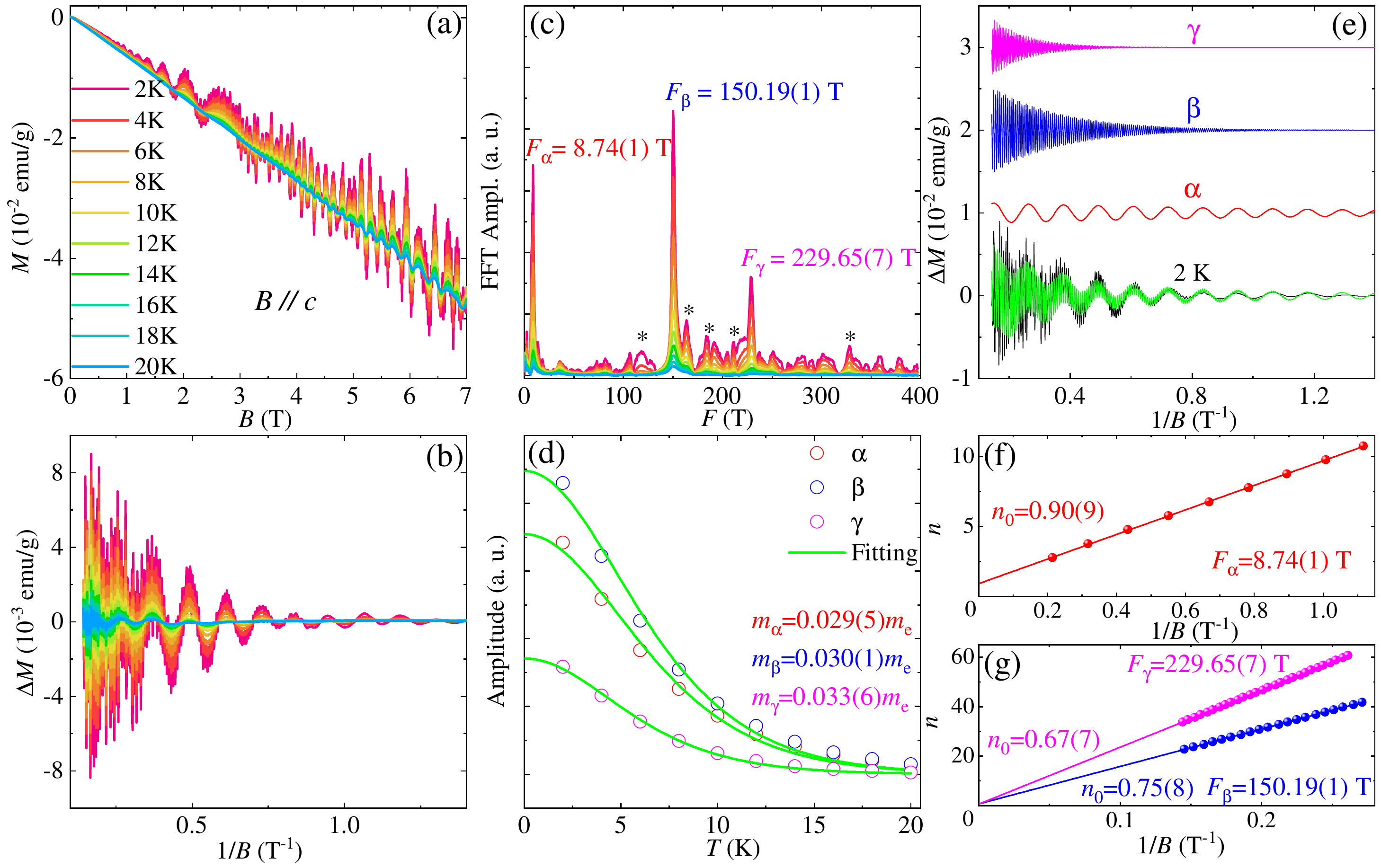}
\begin{center}
\caption{\label{dHvA} (Color online) The dHvA oscillations and nontrivial Berry phase of Cu$_3$Sn. (a) Isothermal magnetization under $\emph{\textbf{B}}\parallel c$ at different temperatures from 2 K to 20 K. (b) The magnetization oscillations after subtracting the polynomial background. (c) The corresponding FFT spectrum. Several small peaks are marked by the asterisks. These small peaks may be derived from the specific orbitals in the Fermi surface, or they can be related to some measurement noises. (d) The FFT amplitude as a function of temperature and the fit to $R_\textmd{T}$ to determine the effective mass. (e) The Lifshitz-Kosevich fit (green line) of the oscillation pattern at 2 K and the extracted single-frequency
oscillatory components. The single-frequency parts are shifted vertically for clarity. The Landau's fan diagram for the identified frequencies, $F_\alpha$ (f), $F_\beta$ and $F_\gamma$ (g).}
\end{center}
\end{figure*}

Having determined the $a$-axis and $b$-axis of the crystals (see the insets of Fig.~\ref{rt}(a) and (d)), we turn to measure the electrical resistivity along these two directions. The temperature ($T$) dependence of electrical resistivity $\rho_{b}$ and $\rho_{a}$ is shown in Fig.~\ref{rt}(a) and (d), respectively. As seen, Cu$_3$Sn displays a high degree of metallicity in the whole temperature range studied and has a remarkably low resistivity, i.e., $\rho_{b}$ = 7.09 $\mu\Omega\cdot$cm ($\rho_{a}$ = 9.03 $\mu\Omega\cdot$cm) at room temperature, falling to 0.11 $\mu\Omega\cdot$cm (0.21 $\mu\Omega\cdot$cm) at 1.8 K, yielding an residual resistivity ratio RRR = $\rho_{b}$(300K)/$\rho_{b}$(1.8K) = 64 ($\rho_{a}$(300K)/$\rho_{a}$(1.8K) = 43). At low temperatures, $\rho_{a}$ and $\rho_{b}$ show a power-law dependence on temperature (see the insets in Fig.~\ref{rt}(a) and (d)), indicating the dominant electron-electron scattering. At high temperatures (e.g., $T$$\geq$50 K), the resistivity shows quasi-linear behaviors in temperature due to dominant electron-phonon scattering, which can be further modelled with the Bloch-Gr\"{u}neisen (BG) formula \cite{BG}:

\begin{align}
\rho(T) = \rho_0 + A(\frac{T}{\Theta_D})^5\int_0^{\frac{\Theta_D}{T}}\frac{x^5}{(e^x-1)(1-e^{-x})}dx,
\end{align}

\noindent where $\rho_0$, $A$ and $\Theta_D$ are the residual resistivity, electron-phonon interaction constant, Debye temperature respectively. The red solid lines in Fig.~\ref{rt}(a) and Fig.~\ref{rt}(d) represent the fits with the above equation. The fitting yields $\rho_0$ = 0.17 $\pm$ 0.01 $\mu\Omega\cdot$cm, $A$ = 22.68 $\pm$ 0.02 $\mu\Omega\cdot$cm, $\Theta_D$ = 235 $\pm$ 2 K for $\rho_b$ and $\rho_0$ = 0.34 $\pm$ 0.01 $\mu\Omega\cdot$cm, $A$ = 26.63 $\pm$ 0.02 $\mu\Omega\cdot$cm, $\Theta_D$ = 220 $\pm$ 2 K for $\rho_a$.

With the magnetic field applied along the $c$ axis, both $\rho_{b}$ and $\rho_{a}$ show nearly identical $T$ dependence with that in zero magnetic field, as shown in the insets of Fig.~\ref{rt}(a) and (d), i.e., with no field-induced resistivity upturns. We also measured the field-dependent MR with the magnetic field applied along the $c$ axis at fixed temperatures below 100 K, defined as $\Delta \rho$/$\rho$(0 T) [$\Delta \rho$ = $\rho$ $-$ $\rho$(0 T)]. As shown in Fig.~\ref{rt}(b), the transverse MR of $\rho_{b}$ reaches only 154\% at 1.8 K and 9 T and varies quasi-linearly in field without saturation. This quasi-linear MR is more clearly seen in $\rho_a$ (Fig.~\ref{rt}(e)). In the same setting, the MR in $\rho_{a}$ is approximately a factor of 4 larger than in $\rho_{b}$ ($\Delta \rho_a/\rho_a$ reaching a magnitude as high as 642\% at 2 K and 9 T), reflecting the anisotropy in the $a$-axis and $b$-axis transport. The nonsaturating quasi-linear MR has been observed in many topological semimetals including WTe$_2$ \cite{Ali-nature}, PtSn$_4$ \cite{PtSn4}, TaPdTe$_5$ \cite{TaPdTe5} and was invoked as evidence for the existence of topological carriers. For topological materials hosting Dirac fermions with a linear energy dispersion, the linear MR could generically arise due to the small quantum limit in magnetic field \cite{PtSn4}.

In the framework of the Boltzmann theory, the Kohler's rule describes the form of MR in magnetic field \cite{Nigel-Kohler} and has been found to be well obeyed in a large number of standard metals, including the single-band and multiband systems \cite{Pippard,TaSe3}. Generally, the Kohler's rule simply dictates $\Delta \rho$/$\rho_0$= $f$($H/\rho_0$) so that the MR at different temperatures will collapse onto a single curve when it is plotted as a function of $H/\rho_0$. Here $\rho_0$ is the zero-field resistivity. The violation of this rule implies either the change of carrier density with temperature or the anisotropic electron scattering $\tau$($k$) that changes the form with temperature~\cite{Nigel-Kohler,extendedKohler}. By the same token, the validity of the Kohler's rule indicates the carrier density does not change significantly with temperature or there is no strange electron scattering of the underlying Fermi surface (FS). The Kohler's rule in Cu$_3$Sn, as plotted in Fig.~\ref{rt}(c) and Fig.~\ref{rt}(f), is overall obeyed, indicating that there is neither strange scattering nor the prominent change of carrier density with temperature. This is conceivable in this Cu$_3$Sn system from the Hall measurement. The Hall resistivity $\rho_{xy}$ as a function of field, measured with the electrical current flowing along the $b$ axis and the magnetic field applied along the $c$ axis, at several fixed temperatures below 200 K, is shown in Fig.~\ref{rt}(g). As seen, $\rho_{xy}$ varies nearly linearly in field. The linear Hall resistivity in a multiband system indicates either the perfectly compensated electrons and holes, or the dominant one type of charge carriers. From the calculations shown below, the electron pockets dominate in Cu$_3$Sn. The negative values of $\rho_{xy}$ in the whole temperature range also indicate the dominance of electron-type carriers, consistent with the calculations. The Hall coefficient $R_{\textmd{H}}$ can be extracted from the linear fitting of $\rho_{xy}$ and was plotted in Fig.~\ref{rt}(h). We find that $R_{\textmd{H}}$ is only weakly temperature dependent. Since electron-type carriers dominate, the back-of-envelope calculation gives its carrier density and the Hall mobility by $n$ = 1/$R_{\textmd{H}}e$ and $\mu_\textmd{H}$ = $R_{\textmd{H}}$/$\rho_0$ at 5 K, yielding $n$ $\sim$ 2.68 $\times$ 10$^{22}$ cm$^{-3}$ and $\mu_\textmd{H}$ $\sim$ 4.91 $\times$ 10$^{2}$ cm$^{2}$ V$^{-1}$s$^{-1}$. Because of this high carrier density (comparable to many other systems in which the Kohler's rule was reported to be valid~\cite{extendedKohler}), it makes the thermally induced changes in the carrier density irrelevant, therefore justifies the validity of the Kohler's rule in this system.

\begin{figure*}[t]
\begin{center}
\includegraphics[width=0.9\textwidth]{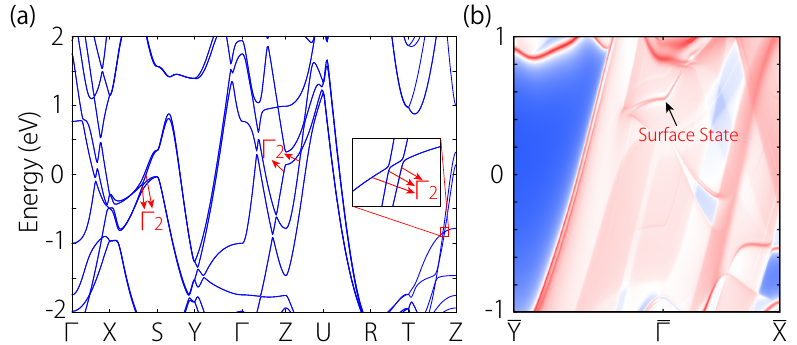}
\caption{\label{band} (a) Calculated band structure of Cu$_{3}$Sn with spin-orbit coupling included. The red arrows indicate the double irreducible representations of energy bands. (b) Surface spectrum for the (001) surface.  The balck arrow indicates nontrivial surface state.}
\end{center}
\end{figure*}

\begin{table*}
\caption{\label{Pa}Parameters obtained from the dHvA oscillations of a Cu$_3$Sn single crystal. The parameters from the Shubnikov-de Haas (SdH) oscillations of the Ag$_3$Sn single crystal are included as a comparison~\cite{ZN-prb}. $F$ is the frequency in the FFT spectra; $A_e$ is the extremal Fermi-surface cross-sectional area calculated from the Onsager relation and $k_F$ is the Fermi wave vector; $m^*$ and $m_e$ are the effective electron mass and the bare electron mass, respectively; $T_\textmd{D}$ is the Dingle temperature; $\tau_q$ is the quantum lifetime.}

\begin{tabular}{l l l l l l l l l l l}
\hline
\hline
dHvA($B\parallel c$)	&  & Cu$_3$Sn	&  &   &Ag$_3$Sn \\
\hline
Parameters  &  $\alpha$   &    $\beta$   &  $\gamma$  &  $\alpha$'   &    $\beta$'   &  $\gamma$'   \\
\hline
$F$(T) &8.74 & 150.19&229.66  &  138.1(9)   &   172.9(9)   &  1566.4(9)    \\
$A_e$(10$^{-3}$ {\AA}$^{-2}$) & 0.83 & 14.31 & 21.88  &  13   &    16   &  149  \\
$k_F$(10$^{-2}$ {\AA}$^{-1}$)& 1.63  &  6.75 & 8.35 &  6.5   &    7.1   &  21.8  \\
$m^*/m_e$ & 0.029& 0.030&  0.033  &  0.05(2)   &    0.04(6)   &  0.09(3)  \\
$T_\textmd{D}$(K) & 1.80 & 9.81&  16.79  &  104.2(3)   &    109.1(4)   &  26.0(8)  \\
$\tau_q$(10$^{-13}$ s) & 6.76 & 1.24&  7.25 &  0.11(7)   &    0.11(1)   &  0.46(6)  \\
\hline
\hline

\end{tabular}
\end{table*}

The isothermal magnetization measured up to 7 T under $\emph{\textbf{B}}\parallel c$ for Cu$_3$Sn is shown in Fig.~\ref{dHvA}(a). The dHvA quantum oscillations can be clearly observed when the magnetic field exceeding 1 T at 2 K, and can sustain up to 20 K. The subtraction of the third-order polynominal background results in the oscillatory components of magnetization at corresponding temperatures, presented in Fig.~\ref{dHvA}(b). From the fast Fourier transform (FFT) analyses of the oscillatory magnetization, we derive three major oscillation frequencies $F_\alpha$ = 8.74(1) T, $F_\beta$=150.19(1) T and $F_\gamma$ = 229.65(7) T, as shown in Fig.~\ref{dHvA}(c). By assuming the circular cross section of the FS along [001], the cross-sectional Fermi area $A_e$ can be calculated from the Onsager relation $A_e$ = $\frac{4e\pi^2F}{h}$ \cite{LK}, where $F$ is the frequency of oscillation. The obtained $A_e$ can further be used to get the Fermi wave vector $k_F$ = $\sqrt{\frac{A_e}{\pi}}$. The corresponding $A_e$ and $k_F$ for different orbitals are listed in Table~\ref{Pa} for concreteness.

In order to obtain the effective mass $m^*$ for each orbit, the oscillatory dHvA magnetization is further analyzed by the Lifshitz-Kosevich (LK) formula \cite{LK}, which takes the Berry phase into account for a Dirac system \cite{Berry}:
\begin{align}
\Delta M \propto - R_T \cdot R_D \cdot R_S \cdot sin[2\pi(\frac{F}{B} + \frac{1}{2} - \frac{\varphi_B}{2\pi} - \delta)],
\end{align}

\noindent where $R_\textmd{T}$ = $\frac{2\pi^2k_\textmd{B}m^* T/\hbar e B}{sinh(2\pi^2k_\textmd{B}m^* T/\hbar e B)}$, $R_\textmd{D}$ = exp($-$2$\pi^2k_\textmd{B}T_\textmd{D}m^*$/$\hbar eB$), and $R_\textmd{S}$ = cos($\pi gm^*$/2$m_e$) are the thermal damping factor, Dingle damping term, and a spin-related damping term, respectively. $T_\textmd{D}$ is the Dingle temperature and $\varphi_\textmd{B}$ is the Berry phase. The oscillation of $\Delta M$ is described by the sine term with the phase factor $\frac{1}{2} - \frac{\varphi_B}{2\pi} - \delta$. The additional phase shift $\delta$ is determined by the dimensionality of the FS, and is equal to 0 for a two-dimensional FS and $\pm$ 1/8 for a 3D FS. The fitting of the temperature-dependent FFT amplitude to the thermal damping factor $R_\textmd{T}$, as shown in Fig.~\ref{dHvA}(d), yields effective masses corresponding to three fundamental frequencies $m_\alpha^*$ = 0.029(5)$m_e$, $m_\beta^*$ = 0.030(1)$m_e$ and $m_\gamma^*$ = 0.033(6)$m_e$ ($m_e$ is the bare electron mass), all of which are extremely small. With the fixed parameters of three fundamental frequencies and corresponding effective masses, the oscillation pattern can be well fitted by using the LK formula directly. The extracted single-frequency oscillatory signals are plotted separately in Fig.~\ref{dHvA}(e), and the LK formula well reproduces the oscillations at 2 K. The quantum relaxation time $\tau_q$ can be obtained from the Dingle temperature by $\tau_q$ = $\hbar$/(2$\pi k_\textmd{B}T_\textmd{D}$). The corresponding values for $T_\textmd{D}$, $\tau_q$ and $m^*$ for different bands are also listed in Table~\ref{Pa}. The phase factors are fitted to be 1.06 ($F_\alpha$), 1.02 ($F_\beta$), and 1.29 ($F_\gamma$), from which the Berry phases $\varphi_B$ are derived to be 0.12$\pi$, 0.37$\pi$ and $-$0.13$\pi$ ($F_\alpha$), 0.04$\pi$, 0.33$\pi$ and $-$0.17$\pi$ ($F_\beta$) and 0.58$\pi$, 1.41$\pi$ and 0.91$\pi$ ($F_\gamma$) for $\delta$ = 0, $-$1/8 and 1/8, respectively. In general, the Berry phase is zero for a parabolic energy dispersion and $\pi$ for a linear energy dispersion. A finite Berry phase, deviating from the exact $\pi$ value, is also possible due to the deviation of the dispersion relation $E$($k$) from an ideal linear dispersion \cite{Berry_deviation}. In addition, the Berry phase can be further evaluated using the Landau level (LL) fan diagram \cite{LL}. The minima of $\Delta M$ should be assigned to a LL index of $n$ $-$ 1/4 \cite{Mao-prl}. The established LL fan diagram is shown in Fig.~\ref{dHvA}(f) and (g). The linear extrapolation in the fan diagram yields three intercepts $n_0$ = 0.91 ($F_\alpha$), $n_0$ = 0.76 ($F_\beta$) and $n_0$ = 0.68 ($F_\gamma$), corresponding to a nontrivial Berry phases of $\varphi_\textmd{B}$ = 2$\pi(0.91 + \delta)$ ($F_\alpha$), $\varphi_\textmd{B}$ = 2$\pi(0.76 + \delta)$ ($F_\beta$) and $\varphi_\textmd{B}$ = 2$\pi(0.68 + \delta)$ ($F_\gamma$). The slopes of the LL fan diagram are 8.78 T, 151.19 T and 229.85 T for the $F_\alpha$, $F_\beta$ and $F_\gamma$ bands, respectively, in excellent agreement with the oscillation frequencies identified through the FFT analysis.

In order to gain more insights into the topological properties of Cu$_3$Sn, we investigate, based on the first-principles calculations, the bulk band structures along the high-symmetry lines as shown in Fig. \ref{band}(a). Here the spin-orbit coupling is considered. One can observe that Cu$_3$Sn exhibits the metal character. Several bands around the Fermi level cross each other, which in fact open small band-gaps owing to the same double irreducible representation ($\Gamma_{2}$). Furthermore, as the system respects the inversion symmetry, $\mathbb{Z}_2$ invariant of this system can be evaluated using the Fu-Kane formula as shown in Ref. \cite{PhysRevB.76.045302}. We find $\mathbb{Z}_2$=1 by calculating the parities for all the occupied bands at 8 time-reversal invariant momenta \cite{PhysRevB.76.045302}, suggesting the strong topological state in Cu$_{3}$Sn. Furthermore, in Fig. \ref{band}(b), we show the results of the surface state along $\bar{X}-\bar{\Gamma}-\bar{Y}$ for the (001) surface by combining the Wannier-basis tight-binding model with the surface Green's function method \cite{MOSTOFI2008685,Sancho_1985,WU2018405}. It is clear to see that there exists nontrivial Dirac-cone-like surface state located at the $\bar{\Gamma}$ point, further supporting the calculation of $\mathbb{Z}_2$=1.

There are four bands crossing the Fermi level $E_F$ and their individual FSs are shown in the SM, which demonstrates strong 3D characteristics. Band 1 and 2 are hole-like while Band 3 and 4 are electron-like. The resultant FSs show strong similarity between two hole(electron)-like pockets and signify the dominance of the electron-like carriers, consistent with the Hall measurements.

\section{Summary}
In summary, we have successfully grown the high-quality single crystals of an IMS Cu$_3$Sn that is isostructural with the topological semimetal Ag$_3$Sn.
The quasi-linear MR and the quantum dHvA oscillations, featuring extremely small effective masses and the possible nonzero Berry phases, suggest that Cu$_3$Sn may be another topologically nontrivial stannide, in common with its analogues Ag$_3$Sn and AuSn$_4$. Our first-principles calculations further corroborate its nontrivial band topology. More experiments, such as the angle dependent dHvA oscillations, ARPES, are highly desirable to provide further evidence for its Fermiology and the putative topological states proposed in this study.

\begin{center}
{\bf ACKNOWLEDGEMENTS}
\end{center}

The authors would like to thank Xiangang Wan for fruitful discussions. This work was supported by the National Natural Science Foundation of China (No. 11974061) and the users with Excellence Project of Hefei Science Center of CAS (2021HSC-UE007). B. Li was supported by the National Natural Science Foundation of China (No. 11674054) and NUPTSF (Nos. NY220038 and NY219087).

\bibliography{Cu3Sn}

\begin{thebibliography}{40}%
\makeatletter
\providecommand \@ifxundefined [1]{%
 \@ifx{#1\undefined}
}%
\providecommand \@ifnum [1]{%
 \ifnum #1\expandafter \@firstoftwo
 \else \expandafter \@secondoftwo
 \fi
}%
\providecommand \@ifx [1]{%
 \ifx #1\expandafter \@firstoftwo
 \else \expandafter \@secondoftwo
 \fi
}%
\providecommand \natexlab [1]{#1}%
\providecommand \enquote  [1]{``#1''}%
\providecommand \bibnamefont  [1]{#1}%
\providecommand \bibfnamefont [1]{#1}%
\providecommand \citenamefont [1]{#1}%
\providecommand \href@noop [0]{\@secondoftwo}%
\providecommand \href [0]{\begingroup \@sanitize@url \@href}%
\providecommand \@href[1]{\@@startlink{#1}\@@href}%
\providecommand \@@href[1]{\endgroup#1\@@endlink}%
\providecommand \@sanitize@url [0]{\catcode `\\12\catcode `\$12\catcode
  `\&12\catcode `\#12\catcode `\^12\catcode `\_12\catcode `\%12\relax}%
\providecommand \@@startlink[1]{}%
\providecommand \@@endlink[0]{}%
\providecommand \url  [0]{\begingroup\@sanitize@url \@url }%
\providecommand \@url [1]{\endgroup\@href {#1}{\urlprefix }}%
\providecommand \urlprefix  [0]{URL }%
\providecommand \Eprint [0]{\href }%
\providecommand \doibase [0]{https://doi.org/}%
\providecommand \selectlanguage [0]{\@gobble}%
\providecommand \bibinfo  [0]{\@secondoftwo}%
\providecommand \bibfield  [0]{\@secondoftwo}%
\providecommand \translation [1]{[#1]}%
\providecommand \BibitemOpen [0]{}%
\providecommand \bibitemStop [0]{}%
\providecommand \bibitemNoStop [0]{.\EOS\space}%
\providecommand \EOS [0]{\spacefactor3000\relax}%
\providecommand \BibitemShut  [1]{\csname bibitem#1\endcsname}%
\let\auto@bib@innerbib\@empty
\bibitem [{\citenamefont {Armitage}\ \emph {et~al.}(2018)\citenamefont
  {Armitage}, \citenamefont {Mele},\ and\ \citenamefont
  {Vishwanath}}]{armitage-review}%
  \BibitemOpen
  \bibfield  {author} {\bibinfo {author} {\bibfnamefont {N.~P.}\ \bibnamefont
  {Armitage}}, \bibinfo {author} {\bibfnamefont {E.~J.}\ \bibnamefont {Mele}},\
  and\ \bibinfo {author} {\bibfnamefont {A.}~\bibnamefont {Vishwanath}},\
  }\bibfield  {title} {\bibinfo {title} {Weyl and \textsc{D}irac semimetals in
  three-dimensional solids},\ }\href
  {https://doi.org/10.1103/RevModPhys.90.015001} {\bibfield  {journal}
  {\bibinfo  {journal} {Rev. Mod. Phys.}\ }\textbf {\bibinfo {volume} {90}},\
  \bibinfo {pages} {015001} (\bibinfo {year} {2018})}\BibitemShut {NoStop}%
\bibitem [{\citenamefont {Lv}\ \emph {et~al.}(2021)\citenamefont {Lv},
  \citenamefont {Qian},\ and\ \citenamefont {Ding}}]{ding-review}%
  \BibitemOpen
  \bibfield  {author} {\bibinfo {author} {\bibfnamefont {B.~Q.}\ \bibnamefont
  {Lv}}, \bibinfo {author} {\bibfnamefont {T.}~\bibnamefont {Qian}},\ and\
  \bibinfo {author} {\bibfnamefont {H.}~\bibnamefont {Ding}},\ }\bibfield
  {title} {\bibinfo {title} {Experimental perspective on three-dimensional
  topological semimetals},\ }\href
  {https://doi.org/https://doi.org/10.1103/RevModPhys.93.025002} {\bibfield
  {journal} {\bibinfo  {journal} {Rev. Mod. Phys.}\ }\textbf {\bibinfo {volume}
  {93}},\ \bibinfo {pages} {025002} (\bibinfo {year} {2021})}\BibitemShut
  {NoStop}%
\bibitem [{\citenamefont {Liang}\ \emph {et~al.}(2015)\citenamefont {Liang},
  \citenamefont {Gibson}, \citenamefont {Ali}, \citenamefont {Liu},
  \citenamefont {Cava},\ and\ \citenamefont {Ong}}]{Ong}%
  \BibitemOpen
  \bibfield  {author} {\bibinfo {author} {\bibfnamefont {T.}~\bibnamefont
  {Liang}}, \bibinfo {author} {\bibfnamefont {Q.}~\bibnamefont {Gibson}},
  \bibinfo {author} {\bibfnamefont {M.~N.}\ \bibnamefont {Ali}}, \bibinfo
  {author} {\bibfnamefont {M.}~\bibnamefont {Liu}}, \bibinfo {author}
  {\bibfnamefont {R.~J.}\ \bibnamefont {Cava}},\ and\ \bibinfo {author}
  {\bibfnamefont {N.~P.}\ \bibnamefont {Ong}},\ }\bibfield  {title} {\bibinfo
  {title} {Ultrahigh mobility and giant magnetoresistance in the dirac
  semimetal {Cd$_3$As$_2$}},\ }\href
  {https://doi.org/https://doi.org/10.1038/nmat4143} {\bibfield  {journal}
  {\bibinfo  {journal} {Nat. Mater}\ }\textbf {\bibinfo {volume} {14}},\
  \bibinfo {pages} {280} (\bibinfo {year} {2015})}\BibitemShut {NoStop}%
\bibitem [{\citenamefont {Parameswaran}\ \emph {et~al.}(2014)\citenamefont
  {Parameswaran}, \citenamefont {Grover}, \citenamefont {Abanin}, \citenamefont
  {Pesin},\ and\ \citenamefont {Vishwanath}}]{Parame}%
  \BibitemOpen
  \bibfield  {author} {\bibinfo {author} {\bibfnamefont {S.~A.}\ \bibnamefont
  {Parameswaran}}, \bibinfo {author} {\bibfnamefont {T.}~\bibnamefont
  {Grover}}, \bibinfo {author} {\bibfnamefont {D.~A.}\ \bibnamefont {Abanin}},
  \bibinfo {author} {\bibfnamefont {D.~A.}\ \bibnamefont {Pesin}},\ and\
  \bibinfo {author} {\bibfnamefont {A.}~\bibnamefont {Vishwanath}},\ }\bibfield
   {title} {\bibinfo {title} {Probing the chiral anomaly with nonlocal
  transport in three-dimensional topological semimetals,},\ }\href
  {https://doi.org/https://doi.org/10.1103/PhysRevX.4.031035} {\bibfield
  {journal} {\bibinfo  {journal} {Phys. Rev. X}\ }\textbf {\bibinfo {volume}
  {4}},\ \bibinfo {pages} {031035} (\bibinfo {year} {2014})}\BibitemShut
  {NoStop}%
\bibitem [{\citenamefont {Huang}\ \emph {et~al.}(2015)\citenamefont {Huang},
  \citenamefont {Zhao}, \citenamefont {Long}, \citenamefont {Wang},
  \citenamefont {Chen}, \citenamefont {Yang}, \citenamefont {Liang},
  \citenamefont {Xue}, \citenamefont {Weng}, \citenamefont {Fang},
  \citenamefont {Dai},\ and\ \citenamefont {Chen}}]{Chen}%
  \BibitemOpen
  \bibfield  {author} {\bibinfo {author} {\bibfnamefont {X.~C.}\ \bibnamefont
  {Huang}}, \bibinfo {author} {\bibfnamefont {L.~X.}\ \bibnamefont {Zhao}},
  \bibinfo {author} {\bibfnamefont {Y.~J.}\ \bibnamefont {Long}}, \bibinfo
  {author} {\bibfnamefont {P.~P.}\ \bibnamefont {Wang}}, \bibinfo {author}
  {\bibfnamefont {D.}~\bibnamefont {Chen}}, \bibinfo {author} {\bibfnamefont
  {Z.~H.}\ \bibnamefont {Yang}}, \bibinfo {author} {\bibfnamefont
  {H.}~\bibnamefont {Liang}}, \bibinfo {author} {\bibfnamefont {M.~Q.}\
  \bibnamefont {Xue}}, \bibinfo {author} {\bibfnamefont {H.~M.}\ \bibnamefont
  {Weng}}, \bibinfo {author} {\bibfnamefont {Z.}~\bibnamefont {Fang}}, \bibinfo
  {author} {\bibfnamefont {X.}~\bibnamefont {Dai}},\ and\ \bibinfo {author}
  {\bibfnamefont {G.~F.}\ \bibnamefont {Chen}},\ }\bibfield  {title} {\bibinfo
  {title} {Observation of the chiral-anomaly-induced negative magnetoresistance
  in {3D} weyl semimetal {TaAs}},\ }\href
  {https://doi.org/https://doi.org/10.1103/PhysRevX.5.031023} {\bibfield
  {journal} {\bibinfo  {journal} {Phys. Rev. X}\ }\textbf {\bibinfo {volume}
  {5}},\ \bibinfo {pages} {031023} (\bibinfo {year} {2015})}\BibitemShut
  {NoStop}%
\bibitem [{\citenamefont {Kumar}\ \emph {et~al.}(2018)\citenamefont {Kumar},
  \citenamefont {Guin}, \citenamefont {Felser},\ and\ \citenamefont
  {Shekhar}}]{PHE-Kumar}%
  \BibitemOpen
  \bibfield  {author} {\bibinfo {author} {\bibfnamefont {N.}~\bibnamefont
  {Kumar}}, \bibinfo {author} {\bibfnamefont {S.~N.}\ \bibnamefont {Guin}},
  \bibinfo {author} {\bibfnamefont {C.}~\bibnamefont {Felser}},\ and\ \bibinfo
  {author} {\bibfnamefont {C.}~\bibnamefont {Shekhar}},\ }\bibfield  {title}
  {\bibinfo {title} {Planar \textsc{H}all effect in the \textsc{W}eyl semimetal
  {GdPtBi}},\ }\href
  {https://doi.org/https://doi.org/10.1103/PhysRevB.98.041103} {\bibfield
  {journal} {\bibinfo  {journal} {Phys. Rev. B}\ }\textbf {\bibinfo {volume}
  {98}},\ \bibinfo {pages} {041103} (\bibinfo {year} {2018})}\BibitemShut
  {NoStop}%
\bibitem [{\citenamefont {He}\ \emph {et~al.}(2014)\citenamefont {He},
  \citenamefont {Hong}, \citenamefont {Dong}, \citenamefont {Pan},
  \citenamefont {Zhang}, \citenamefont {Zhang},\ and\ \citenamefont
  {Li}}]{He-prl}%
  \BibitemOpen
  \bibfield  {author} {\bibinfo {author} {\bibfnamefont {L.~P.}\ \bibnamefont
  {He}}, \bibinfo {author} {\bibfnamefont {X.~C.}\ \bibnamefont {Hong}},
  \bibinfo {author} {\bibfnamefont {J.~K.}\ \bibnamefont {Dong}}, \bibinfo
  {author} {\bibfnamefont {J.}~\bibnamefont {Pan}}, \bibinfo {author}
  {\bibfnamefont {Z.}~\bibnamefont {Zhang}}, \bibinfo {author} {\bibfnamefont
  {J.}~\bibnamefont {Zhang}},\ and\ \bibinfo {author} {\bibfnamefont {S.~Y.}\
  \bibnamefont {Li}},\ }\bibfield  {title} {\bibinfo {title} {Quantum transport
  evidence for the three-dimensional dirac demimetal phase in {Cd$_3$As$_2$}},\
  }\href {https://doi.org/https://doi.org/10.1103/PhysRevLett.113.246402}
  {\bibfield  {journal} {\bibinfo  {journal} {Phys. Rev. Lett.}\ }\textbf
  {\bibinfo {volume} {113}},\ \bibinfo {pages} {246402} (\bibinfo {year}
  {2014})}\BibitemShut {NoStop}%
\bibitem [{\citenamefont {Xiang}\ \emph {et~al.}(2015)\citenamefont {Xiang},
  \citenamefont {Zhao}, \citenamefont {Jin}, \citenamefont {Shang},
  \citenamefont {Ma}, \citenamefont {Ye}, \citenamefont {Lei}, \citenamefont
  {Wu}, \citenamefont {Xia},\ and\ \citenamefont {Chen}}]{CXH-prl}%
  \BibitemOpen
  \bibfield  {author} {\bibinfo {author} {\bibfnamefont {Z.~J.}\ \bibnamefont
  {Xiang}}, \bibinfo {author} {\bibfnamefont {D.}~\bibnamefont {Zhao}},
  \bibinfo {author} {\bibfnamefont {Z.}~\bibnamefont {Jin}}, \bibinfo {author}
  {\bibfnamefont {C.}~\bibnamefont {Shang}}, \bibinfo {author} {\bibfnamefont
  {L.~K.}\ \bibnamefont {Ma}}, \bibinfo {author} {\bibfnamefont {G.~J.}\
  \bibnamefont {Ye}}, \bibinfo {author} {\bibfnamefont {B.}~\bibnamefont
  {Lei}}, \bibinfo {author} {\bibfnamefont {T.}~\bibnamefont {Wu}}, \bibinfo
  {author} {\bibfnamefont {Z.~C.}\ \bibnamefont {Xia}},\ and\ \bibinfo {author}
  {\bibfnamefont {X.~H.}\ \bibnamefont {Chen}},\ }\bibfield  {title} {\bibinfo
  {title} {Angular-dependent phase factor of shubnikov--de haas oscillations in
  the dirac semimetal ${\mathrm{cd}}_{3}{\mathrm{as}}_{2}$},\ }\href
  {https://doi.org/10.1103/PhysRevLett.115.226401} {\bibfield  {journal}
  {\bibinfo  {journal} {Phys. Rev. Lett.}\ }\textbf {\bibinfo {volume} {115}},\
  \bibinfo {pages} {226401} (\bibinfo {year} {2015})}\BibitemShut {NoStop}%
\bibitem [{\citenamefont {Liu}\ \emph {et~al.}(2018)\citenamefont {Liu},
  \citenamefont {Sun}, \citenamefont {Kumar}, \citenamefont {Muechler},
  \citenamefont {Sun},\ and\ \citenamefont {et~al.}}]{Liu-np}%
  \BibitemOpen
  \bibfield  {author} {\bibinfo {author} {\bibfnamefont {E.~K.}\ \bibnamefont
  {Liu}}, \bibinfo {author} {\bibfnamefont {Y.}~\bibnamefont {Sun}}, \bibinfo
  {author} {\bibfnamefont {N.}~\bibnamefont {Kumar}}, \bibinfo {author}
  {\bibfnamefont {L.}~\bibnamefont {Muechler}}, \bibinfo {author}
  {\bibfnamefont {A.~L.}\ \bibnamefont {Sun}},\ and\ \bibinfo {author}
  {\bibfnamefont {L.~J.}\ \bibnamefont {et~al.}},\ }\bibfield  {title}
  {\bibinfo {title} {Giant anomalous hall effect in a ferromagnetic
  kagome-lattice semimetal},\ }\href
  {https://doi.org/https://doi.org/10.1038/s41567-018-0234-5} {\bibfield
  {journal} {\bibinfo  {journal} {Nature Phys.}\ }\textbf {\bibinfo {volume}
  {14}},\ \bibinfo {pages} {1125} (\bibinfo {year} {2018})}\BibitemShut
  {NoStop}%
\bibitem [{\citenamefont {Jo}\ \emph {et~al.}(2017)\citenamefont {Jo},
  \citenamefont {Wu}, \citenamefont {Wang}, \citenamefont {Orth}, \citenamefont
  {Downing}, \citenamefont {Manni}, \citenamefont {Mou}, \citenamefont
  {Johnson}, \citenamefont {Kaminski}, \citenamefont {Bud'ko},\ and\
  \citenamefont {Canfield}}]{PtSn4_MR}%
  \BibitemOpen
  \bibfield  {author} {\bibinfo {author} {\bibfnamefont {N.~H.}\ \bibnamefont
  {Jo}}, \bibinfo {author} {\bibfnamefont {Y.}~\bibnamefont {Wu}}, \bibinfo
  {author} {\bibfnamefont {L.}~\bibnamefont {Wang}}, \bibinfo {author}
  {\bibfnamefont {P.~P.}\ \bibnamefont {Orth}}, \bibinfo {author}
  {\bibfnamefont {S.~S.}\ \bibnamefont {Downing}}, \bibinfo {author}
  {\bibfnamefont {S.}~\bibnamefont {Manni}}, \bibinfo {author} {\bibfnamefont
  {D.}~\bibnamefont {Mou}}, \bibinfo {author} {\bibfnamefont {D.~D.}\
  \bibnamefont {Johnson}}, \bibinfo {author} {\bibfnamefont {A.}~\bibnamefont
  {Kaminski}}, \bibinfo {author} {\bibfnamefont {S.~L.}\ \bibnamefont
  {Bud'ko}},\ and\ \bibinfo {author} {\bibfnamefont {P.~C.}\ \bibnamefont
  {Canfield}},\ }\bibfield  {title} {\bibinfo {title} {Extremely large
  magnetoresistance and kohler's rule in {PdSn$_4$}: A complete study of
  thermodynamic, transport, and band-structure properties},\ }\href
  {https://doi.org/https://doi.org/10.1103/PhysRevB.96.165145} {\bibfield
  {journal} {\bibinfo  {journal} {Phys. Rev. B}\ }\textbf {\bibinfo {volume}
  {96}},\ \bibinfo {pages} {165145} (\bibinfo {year} {2017})}\BibitemShut
  {NoStop}%
\bibitem [{\citenamefont {Wu}\ \emph {et~al.}(2016)\citenamefont {Wu},
  \citenamefont {Wang}, \citenamefont {Mun}, \citenamefont {Johnson},
  \citenamefont {Mou}, \citenamefont {Huang}, \citenamefont {Lee},
  \citenamefont {Bud'ko}, \citenamefont {Canfield},\ and\ \citenamefont
  {Kaminski}}]{PtSn4_ARPES}%
  \BibitemOpen
  \bibfield  {author} {\bibinfo {author} {\bibfnamefont {Y.}~\bibnamefont
  {Wu}}, \bibinfo {author} {\bibfnamefont {L.}~\bibnamefont {Wang}}, \bibinfo
  {author} {\bibfnamefont {E.}~\bibnamefont {Mun}}, \bibinfo {author}
  {\bibfnamefont {D.~D.}\ \bibnamefont {Johnson}}, \bibinfo {author}
  {\bibfnamefont {D.}~\bibnamefont {Mou}}, \bibinfo {author} {\bibfnamefont
  {L.}~\bibnamefont {Huang}}, \bibinfo {author} {\bibfnamefont
  {Y.}~\bibnamefont {Lee}}, \bibinfo {author} {\bibfnamefont {S.~L.}\
  \bibnamefont {Bud'ko}}, \bibinfo {author} {\bibfnamefont {P.~C.}\
  \bibnamefont {Canfield}},\ and\ \bibinfo {author} {\bibfnamefont
  {A.}~\bibnamefont {Kaminski}},\ }\bibfield  {title} {\bibinfo {title} {Dirac
  node arcs in {PtSn$_4$}},\ }\href {https://doi.org/10.1038/NPHYS3712}
  {\bibfield  {journal} {\bibinfo  {journal} {Nat. Phys.}\ }\textbf {\bibinfo
  {volume} {12}},\ \bibinfo {pages} {667} (\bibinfo {year} {2016})}\BibitemShut
  {NoStop}%
\bibitem [{\citenamefont {Xu}\ \emph {et~al.}(2017)\citenamefont {Xu},
  \citenamefont {Zhou}, \citenamefont {Sankar}, \citenamefont {Xing},
  \citenamefont {Shi}, \citenamefont {Han}, \citenamefont {Qian}, \citenamefont
  {Wang}, \citenamefont {Zhu}, \citenamefont {Zhang}, \citenamefont {Bangura},
  \citenamefont {Hussey},\ and\ \citenamefont {Xu}}]{PtSn4}%
  \BibitemOpen
  \bibfield  {author} {\bibinfo {author} {\bibfnamefont {C.~Q.}\ \bibnamefont
  {Xu}}, \bibinfo {author} {\bibfnamefont {W.}~\bibnamefont {Zhou}}, \bibinfo
  {author} {\bibfnamefont {R.}~\bibnamefont {Sankar}}, \bibinfo {author}
  {\bibfnamefont {X.~Z.}\ \bibnamefont {Xing}}, \bibinfo {author}
  {\bibfnamefont {Z.~X.}\ \bibnamefont {Shi}}, \bibinfo {author} {\bibfnamefont
  {Z.~D.}\ \bibnamefont {Han}}, \bibinfo {author} {\bibfnamefont
  {B.}~\bibnamefont {Qian}}, \bibinfo {author} {\bibfnamefont {J.~H.}\
  \bibnamefont {Wang}}, \bibinfo {author} {\bibfnamefont {Z.~W.}\ \bibnamefont
  {Zhu}}, \bibinfo {author} {\bibfnamefont {J.~L.}\ \bibnamefont {Zhang}},
  \bibinfo {author} {\bibfnamefont {A.~F.}\ \bibnamefont {Bangura}}, \bibinfo
  {author} {\bibfnamefont {N.~E.}\ \bibnamefont {Hussey}},\ and\ \bibinfo
  {author} {\bibfnamefont {X.}~\bibnamefont {Xu}},\ }\bibfield  {title}
  {\bibinfo {title} {Enhanced electron correlations in the binary stannide
  pdsn$_4$: A homologue of the dirac nodal arc semimetal {PtSn$_4$}},\ }\href
  {https://doi.org/https://doi.org/10.1103/PhysRevMaterials.1.064201}
  {\bibfield  {journal} {\bibinfo  {journal} {Phys. Rev. Mater}\ }\textbf
  {\bibinfo {volume} {1}},\ \bibinfo {pages} {064201} (\bibinfo {year}
  {2017})}\BibitemShut {NoStop}%
\bibitem [{\citenamefont {Shen}\ \emph {et~al.}(2020)\citenamefont {Shen},
  \citenamefont {Kuo}, \citenamefont {Yang}, \citenamefont {Chen},
  \citenamefont {Lue},\ and\ \citenamefont {Wang}}]{AuSn4}%
  \BibitemOpen
  \bibfield  {author} {\bibinfo {author} {\bibfnamefont {D.}~\bibnamefont
  {Shen}}, \bibinfo {author} {\bibfnamefont {C.}~\bibnamefont {Kuo}}, \bibinfo
  {author} {\bibfnamefont {T.}~\bibnamefont {Yang}}, \bibinfo {author}
  {\bibfnamefont {I.}~\bibnamefont {Chen}}, \bibinfo {author} {\bibfnamefont
  {C.}~\bibnamefont {Lue}},\ and\ \bibinfo {author} {\bibfnamefont {L.~M.}\
  \bibnamefont {Wang}},\ }\bibfield  {title} {\bibinfo {title} {Two-dimensional
  superconductivity and magnetotransport from topological surface states in
  {AuSn$_4$} semimetal},\ }\href
  {https://doi.org/https://doi.org/10.1038/s43246-020-00060-8} {\bibfield
  {journal} {\bibinfo  {journal} {Commun. Mater.}\ }\textbf {\bibinfo {volume}
  {1}},\ \bibinfo {pages} {56} (\bibinfo {year} {2020})}\BibitemShut {NoStop}%
\bibitem [{\citenamefont {Siddiquee}\ \emph {et~al.}(2022)\citenamefont
  {Siddiquee}, \citenamefont {Munir}, \citenamefont {Dissanayake},
  \citenamefont {Vaidya}, \citenamefont {Nickle}, \citenamefont {Del~Barco},
  \citenamefont {Lamura}, \citenamefont {Baines}, \citenamefont {Cahen},
  \citenamefont {H\'erold}, \citenamefont {Gentile}, \citenamefont {Shiroka},\
  and\ \citenamefont {Nakajima}}]{CaSn3}%
  \BibitemOpen
  \bibfield  {author} {\bibinfo {author} {\bibfnamefont {H.}~\bibnamefont
  {Siddiquee}}, \bibinfo {author} {\bibfnamefont {R.}~\bibnamefont {Munir}},
  \bibinfo {author} {\bibfnamefont {C.}~\bibnamefont {Dissanayake}}, \bibinfo
  {author} {\bibfnamefont {P.}~\bibnamefont {Vaidya}}, \bibinfo {author}
  {\bibfnamefont {C.}~\bibnamefont {Nickle}}, \bibinfo {author} {\bibfnamefont
  {E.}~\bibnamefont {Del~Barco}}, \bibinfo {author} {\bibfnamefont
  {G.}~\bibnamefont {Lamura}}, \bibinfo {author} {\bibfnamefont
  {C.}~\bibnamefont {Baines}}, \bibinfo {author} {\bibfnamefont
  {S.}~\bibnamefont {Cahen}}, \bibinfo {author} {\bibfnamefont
  {C.}~\bibnamefont {H\'erold}}, \bibinfo {author} {\bibfnamefont
  {P.}~\bibnamefont {Gentile}}, \bibinfo {author} {\bibfnamefont
  {T.}~\bibnamefont {Shiroka}},\ and\ \bibinfo {author} {\bibfnamefont
  {Y.}~\bibnamefont {Nakajima}},\ }\bibfield  {title} {\bibinfo {title}
  {Nematic superconductivity in the topological semimetal
  $\mathrm{Ca}{\mathrm{sn}}_{3}$},\ }\href
  {https://doi.org/10.1103/PhysRevB.105.094508} {\bibfield  {journal} {\bibinfo
   {journal} {Phys. Rev. B}\ }\textbf {\bibinfo {volume} {105}},\ \bibinfo
  {pages} {094508} (\bibinfo {year} {2022})}\BibitemShut {NoStop}%
\bibitem [{\citenamefont {Zhou}\ \emph {et~al.}(2020)\citenamefont {Zhou},
  \citenamefont {Sun}, \citenamefont {Xu}, \citenamefont {Xi}, \citenamefont
  {Wang},\ and\ \citenamefont {et~al.}}]{ZN-prb}%
  \BibitemOpen
  \bibfield  {author} {\bibinfo {author} {\bibfnamefont {N.}~\bibnamefont
  {Zhou}}, \bibinfo {author} {\bibfnamefont {Y.}~\bibnamefont {Sun}}, \bibinfo
  {author} {\bibfnamefont {C.~Q.}\ \bibnamefont {Xu}}, \bibinfo {author}
  {\bibfnamefont {C.~Y.}\ \bibnamefont {Xi}}, \bibinfo {author} {\bibfnamefont
  {Z.~S.}\ \bibnamefont {Wang}},\ and\ \bibinfo {author} {\bibfnamefont
  {B.~L.}\ \bibnamefont {et~al.}},\ }\bibfield  {title} {\bibinfo {title}
  {Quantum oscillations and anomalous angle-dependent magnetoresistance in the
  topological candidate {Ag$_3$Sn}},\ }\href
  {https://doi.org/https://doi.org/10.1103/physrevb.101.245102} {\bibfield
  {journal} {\bibinfo  {journal} {Phys. Rev. B}\ }\textbf {\bibinfo {volume}
  {101}},\ \bibinfo {pages} {245102} (\bibinfo {year} {2020})}\BibitemShut
  {NoStop}%
\bibitem [{\citenamefont {Momma}\ and\ \citenamefont {Izumi}(2008)}]{VESTA}%
  \BibitemOpen
  \bibfield  {author} {\bibinfo {author} {\bibfnamefont {K.}~\bibnamefont
  {Momma}}\ and\ \bibinfo {author} {\bibfnamefont {F.}~\bibnamefont {Izumi}},\
  }\bibfield  {title} {\bibinfo {title} {Vesta: a three-dimensional
  visualization system for electronic and structural analysis},\ }\href
  {https://doi.org/https://dx.doi.org/10.1107/S0021889808012016} {\bibfield
  {journal} {\bibinfo  {journal} {J. Appl. Crystallogr.}\ }\textbf {\bibinfo
  {volume} {41}},\ \bibinfo {pages} {653} (\bibinfo {year} {2008})}\BibitemShut
  {NoStop}%
\bibitem [{\citenamefont {Schwarz}\ \emph {et~al.}(2002)\citenamefont
  {Schwarz}, \citenamefont {Blaha},\ and\ \citenamefont {Madsen}}]{Wien2k}%
  \BibitemOpen
  \bibfield  {author} {\bibinfo {author} {\bibfnamefont {K.}~\bibnamefont
  {Schwarz}}, \bibinfo {author} {\bibfnamefont {P.}~\bibnamefont {Blaha}},\
  and\ \bibinfo {author} {\bibfnamefont {G.~K.~H.}\ \bibnamefont {Madsen}},\
  }\bibfield  {title} {\bibinfo {title} {Electronic structure calculations of
  solids using the {WIEN2k} package for material sciences},\ }\href
  {https://doi.org/10.1016/S0010-4655(02)00206-0} {\bibfield  {journal}
  {\bibinfo  {journal} {Comput. Phys. Commun.}\ }\textbf {\bibinfo {volume}
  {147(1-2)}},\ \bibinfo {pages} {71} (\bibinfo {year} {2002})}\BibitemShut
  {NoStop}%
\bibitem [{\citenamefont {Wu}\ and\ \citenamefont {Cohen}(2006)}]{GGA}%
  \BibitemOpen
  \bibfield  {author} {\bibinfo {author} {\bibfnamefont {Z.}~\bibnamefont
  {Wu}}\ and\ \bibinfo {author} {\bibfnamefont {R.~E.}\ \bibnamefont {Cohen}},\
  }\bibfield  {title} {\bibinfo {title} {More accurate generalized gradient
  approximation for solids},\ }\href
  {https://doi.org/https://doi.org/10.1103/PhysRevB.73.235116} {\bibfield
  {journal} {\bibinfo  {journal} {Phys. Rev. B}\ }\textbf {\bibinfo {volume}
  {73}},\ \bibinfo {pages} {235116} (\bibinfo {year} {2006})}\BibitemShut
  {NoStop}%
\bibitem [{\citenamefont {Mostofi}\ \emph {et~al.}(2014)\citenamefont
  {Mostofi}, \citenamefont {Yates}, \citenamefont {Pizzi}, \citenamefont {Lee},
  \citenamefont {Souza}, \citenamefont {Vanderbilt},\ and\ \citenamefont
  {Marzari}}]{wannier1}%
  \BibitemOpen
  \bibfield  {author} {\bibinfo {author} {\bibfnamefont {A.~A.}\ \bibnamefont
  {Mostofi}}, \bibinfo {author} {\bibfnamefont {J.~R.}\ \bibnamefont {Yates}},
  \bibinfo {author} {\bibfnamefont {G.}~\bibnamefont {Pizzi}}, \bibinfo
  {author} {\bibfnamefont {Y.~S.}\ \bibnamefont {Lee}}, \bibinfo {author}
  {\bibfnamefont {I.}~\bibnamefont {Souza}}, \bibinfo {author} {\bibfnamefont
  {D.}~\bibnamefont {Vanderbilt}},\ and\ \bibinfo {author} {\bibfnamefont
  {N.}~\bibnamefont {Marzari}},\ }\bibfield  {title} {\bibinfo {title} {An
  updated version of wannier90: A tool for obtaining maximally-localised
  wannier functions},\ }\href
  {https://doi.org/https://doi.org/10.1016/j.cpc.2014.05.003} {\bibfield
  {journal} {\bibinfo  {journal} {Comput. Phys. Commun.}\ }\textbf {\bibinfo
  {volume} {185}},\ \bibinfo {pages} {2309} (\bibinfo {year}
  {2014})}\BibitemShut {NoStop}%
\bibitem [{\citenamefont {Wu}\ \emph {et~al.}(2018{\natexlab{a}})\citenamefont
  {Wu}, \citenamefont {Zhang}, \citenamefont {Song}, \citenamefont {Troyer},\
  and\ \citenamefont {Soluyanov}}]{wannier2}%
  \BibitemOpen
  \bibfield  {author} {\bibinfo {author} {\bibfnamefont {Q.~S.}\ \bibnamefont
  {Wu}}, \bibinfo {author} {\bibfnamefont {S.~N.}\ \bibnamefont {Zhang}},
  \bibinfo {author} {\bibfnamefont {H.~F.}\ \bibnamefont {Song}}, \bibinfo
  {author} {\bibfnamefont {M.}~\bibnamefont {Troyer}},\ and\ \bibinfo {author}
  {\bibfnamefont {A.}~\bibnamefont {Soluyanov}},\ }\bibfield  {title} {\bibinfo
  {title} {Wanniertools: An open-source software package for novel topological
  materials},\ }\href
  {https://doi.org/https://doi.org/10.1016/j.cpc.2017.09.033} {\bibfield
  {journal} {\bibinfo  {journal} {Comput. Phys. Commun.}\ }\textbf {\bibinfo
  {volume} {224}},\ \bibinfo {pages} {405} (\bibinfo {year}
  {2018}{\natexlab{a}})}\BibitemShut {NoStop}%
\bibitem [{\citenamefont {Burkhardt}\ \emph {et~al.}(1959)\citenamefont
  {Burkhardt}, \citenamefont {Schubert},\ and\ \citenamefont
  {Metallkd}}]{Cu3Sn-59}%
  \BibitemOpen
  \bibfield  {author} {\bibinfo {author} {\bibfnamefont {W.}~\bibnamefont
  {Burkhardt}}, \bibinfo {author} {\bibfnamefont {K.}~\bibnamefont
  {Schubert}},\ and\ \bibinfo {author} {\bibfnamefont {Z.}~\bibnamefont
  {Metallkd}},\ }\bibfield  {title} {\bibinfo {title} {\"{U}ber messingartige
  phasen mit a3-verwandter struktur},\ }\href
  {https://doi.org/10.1515/ijmr-1959-500802} {\bibfield  {journal} {\bibinfo
  {journal} {Phys. Commun.}\ }\textbf {\bibinfo {volume} {50}},\ \bibinfo
  {pages} {442} (\bibinfo {year} {1959})}\BibitemShut {NoStop}%
\bibitem [{\citenamefont {Hendus}\ and\ \citenamefont
  {Kn\"{o}dler}(1956)}]{Cu3Sn-216}%
  \BibitemOpen
  \bibfield  {author} {\bibinfo {author} {\bibfnamefont {H.}~\bibnamefont
  {Hendus}}\ and\ \bibinfo {author} {\bibfnamefont {H.}~\bibnamefont
  {Kn\"{o}dler}},\ }\bibfield  {title} {\bibinfo {title} {Die \"{U}berstruktur
  der $\gamma$-hochtemperaturphase im system kupfer-zinn},\ }\href
  {https://doi.org/https://doi.org/10.1107/S0365110X56002990} {\bibfield
  {journal} {\bibinfo  {journal} {Acta Cryst}\ }\textbf {\bibinfo {volume}
  {9}},\ \bibinfo {pages} {1036} (\bibinfo {year} {1956})}\BibitemShut
  {NoStop}%
\bibitem [{\citenamefont {Kn\"{o}dler}(1957)}]{Cu3Sn-25}%
  \BibitemOpen
  \bibfield  {author} {\bibinfo {author} {\bibfnamefont {H.}~\bibnamefont
  {Kn\"{o}dler}},\ }\bibfield  {title} {\bibinfo {title} {Der strukturelle
  zusammenhang zwischen $\gamma$- und $\epsilon$-phase im system kupfer-zinn},\
  }\href {https://doi.org/https://doi.org/10.1107/S0365110X57000225} {\bibfield
   {journal} {\bibinfo  {journal} {Acta Cryst}\ }\textbf {\bibinfo {volume}
  {10}},\ \bibinfo {pages} {86} (\bibinfo {year} {1957})}\BibitemShut {NoStop}%
\bibitem [{\citenamefont {Sang}\ \emph {et~al.}(2009)\citenamefont {Sang},
  \citenamefont {Du},\ and\ \citenamefont {Ye}}]{Cu3Sn-194}%
  \BibitemOpen
  \bibfield  {author} {\bibinfo {author} {\bibfnamefont {X.~H.}\ \bibnamefont
  {Sang}}, \bibinfo {author} {\bibfnamefont {K.}~\bibnamefont {Du}},\ and\
  \bibinfo {author} {\bibfnamefont {H.~Q.}\ \bibnamefont {Ye}},\ }\bibfield
  {title} {\bibinfo {title} {An ordered structure of cu$_3$sn in cu-sn alloy
  investigated by transmission electron microscopy},\ }\href
  {https://doi.org/https://doi.org/10.1016/j.jallcom.2008.01.107} {\bibfield
  {journal} {\bibinfo  {journal} {Journal of Alloys Compd.}\ }\textbf {\bibinfo
  {volume} {469}},\ \bibinfo {pages} {129} (\bibinfo {year}
  {2009})}\BibitemShut {NoStop}%
\bibitem [{\citenamefont {Cvijovi\'{c}}(2011)}]{BG}%
  \BibitemOpen
  \bibfield  {author} {\bibinfo {author} {\bibfnamefont {D.}~\bibnamefont
  {Cvijovi\'{c}}},\ }\bibfield  {title} {\bibinfo {title} {The bloch-gruneisen
  function of arbitrary order and its series representations},\ }\href
  {https://doi.org/https://doi.org/10.1007/s11232-011-0003-4} {\bibfield
  {journal} {\bibinfo  {journal} {Theor. Math. Phys.}\ }\textbf {\bibinfo
  {volume} {166}},\ \bibinfo {pages} {37} (\bibinfo {year} {2011})}\BibitemShut
  {NoStop}%
\bibitem [{\citenamefont {Ali}\ \emph {et~al.}(2014)\citenamefont {Ali},
  \citenamefont {Xiong}, \citenamefont {Flynn}, \citenamefont {Tao},
  \citenamefont {Gibson}, \citenamefont {Schoop}, \citenamefont {Liang},
  \citenamefont {Haldolaarachchige}, \citenamefont {Hirschberger},
  \citenamefont {Ong},\ and\ \citenamefont {Cava}}]{Ali-nature}%
  \BibitemOpen
  \bibfield  {author} {\bibinfo {author} {\bibfnamefont {M.~N.}\ \bibnamefont
  {Ali}}, \bibinfo {author} {\bibfnamefont {J.}~\bibnamefont {Xiong}}, \bibinfo
  {author} {\bibfnamefont {S.}~\bibnamefont {Flynn}}, \bibinfo {author}
  {\bibfnamefont {J.}~\bibnamefont {Tao}}, \bibinfo {author} {\bibfnamefont
  {Q.~D.}\ \bibnamefont {Gibson}}, \bibinfo {author} {\bibfnamefont {L.~M.}\
  \bibnamefont {Schoop}}, \bibinfo {author} {\bibfnamefont {T.}~\bibnamefont
  {Liang}}, \bibinfo {author} {\bibfnamefont {N.}~\bibnamefont
  {Haldolaarachchige}}, \bibinfo {author} {\bibfnamefont {M.}~\bibnamefont
  {Hirschberger}}, \bibinfo {author} {\bibfnamefont {N.~P.}\ \bibnamefont
  {Ong}},\ and\ \bibinfo {author} {\bibfnamefont {R.~J.}\ \bibnamefont
  {Cava}},\ }\bibfield  {title} {\bibinfo {title} {Large, non-saturating
  magnetoresistance in {WTe$_2$}},\ }\href
  {https://doi.org/https://doi.org/10.1038/nature13763} {\bibfield  {journal}
  {\bibinfo  {journal} {Nature (London)}\ }\textbf {\bibinfo {volume} {514}},\
  \bibinfo {pages} {205} (\bibinfo {year} {2014})}\BibitemShut {NoStop}%
\bibitem [{\citenamefont {Jiao}\ \emph {et~al.}(2020)\citenamefont {Jiao},
  \citenamefont {Xie}, \citenamefont {Liu}, \citenamefont {Xu}, \citenamefont
  {Li},\ and\ \citenamefont {et~al.}}]{TaPdTe5}%
  \BibitemOpen
  \bibfield  {author} {\bibinfo {author} {\bibfnamefont {W.~H.}\ \bibnamefont
  {Jiao}}, \bibinfo {author} {\bibfnamefont {X.~M.}\ \bibnamefont {Xie}},
  \bibinfo {author} {\bibfnamefont {Y.}~\bibnamefont {Liu}}, \bibinfo {author}
  {\bibfnamefont {X.}~\bibnamefont {Xu}}, \bibinfo {author} {\bibfnamefont
  {B.}~\bibnamefont {Li}},\ and\ \bibinfo {author} {\bibfnamefont {C.~Q.~X.}\
  \bibnamefont {et~al.}},\ }\bibfield  {title} {\bibinfo {title} {Topological
  \textsc{D}irac states in a layered telluride {TaPdTe$_5$} with
  quasi-one-dimensional {PdTe$_2$} chains,},\ }\href
  {https://doi.org/https://doi.org/10.1103/PhysRevB.102.075141} {\bibfield
  {journal} {\bibinfo  {journal} {Phys. Rev. B}\ }\textbf {\bibinfo {volume}
  {102}},\ \bibinfo {pages} {075141} (\bibinfo {year} {2020})}\BibitemShut
  {NoStop}%
\bibitem [{\citenamefont {Ayres}\ \emph {et~al.}(2021)\citenamefont {Ayres},
  \citenamefont {Berben}, \citenamefont {\v{C}ulo}, \citenamefont {Hsu},
  \citenamefont {van Heumen},\ and\ \citenamefont {et~al.}}]{Nigel-Kohler}%
  \BibitemOpen
  \bibfield  {author} {\bibinfo {author} {\bibfnamefont {J.}~\bibnamefont
  {Ayres}}, \bibinfo {author} {\bibfnamefont {M.}~\bibnamefont {Berben}},
  \bibinfo {author} {\bibfnamefont {M.}~\bibnamefont {\v{C}ulo}}, \bibinfo
  {author} {\bibfnamefont {Y.~T.}\ \bibnamefont {Hsu}}, \bibinfo {author}
  {\bibfnamefont {E.}~\bibnamefont {van Heumen}},\ and\ \bibinfo {author}
  {\bibfnamefont {Y.~H.}\ \bibnamefont {et~al.}},\ }\bibfield  {title}
  {\bibinfo {title} {Incoherent transport across the strange-metal regime of
  overdoped cuprates},\ }\href
  {https://doi.org/https://doi.org/10.1038/s41586-021-03622-z} {\bibfield
  {journal} {\bibinfo  {journal} {Nature}\ }\textbf {\bibinfo {volume} {595}},\
  \bibinfo {pages} {661} (\bibinfo {year} {2021})}\BibitemShut {NoStop}%
\bibitem [{\citenamefont {Pippard}(1989)}]{Pippard}%
  \BibitemOpen
  \bibfield  {author} {\bibinfo {author} {\bibfnamefont {A.~B.}\ \bibnamefont
  {Pippard}},\ }\bibfield  {title} {\bibinfo {title} {Magnetoresistance in
  metals},\ }\href
  {https://doi.org/https://doi.org/10.1126/science.245.4920.874.a} {\bibfield
  {journal} {\bibinfo  {journal} {Cambridge University, Cambridge, England}\
  }\textbf {\bibinfo {volume} {102}},\ \bibinfo {pages} {075141} (\bibinfo
  {year} {1989})}\BibitemShut {NoStop}%
\bibitem [{\citenamefont {Saleheen}\ \emph {et~al.}(2020)\citenamefont
  {Saleheen}, \citenamefont {Chapai}, \citenamefont {Xing}, \citenamefont
  {Nepal}, \citenamefont {Gong}, \citenamefont {Gui}, \citenamefont {Xie},
  \citenamefont {Young}, \citenamefont {Plummer},\ and\ \citenamefont
  {Jin}}]{TaSe3}%
  \BibitemOpen
  \bibfield  {author} {\bibinfo {author} {\bibfnamefont {A.~I.~U.}\
  \bibnamefont {Saleheen}}, \bibinfo {author} {\bibfnamefont {R.}~\bibnamefont
  {Chapai}}, \bibinfo {author} {\bibfnamefont {L.~Y.}\ \bibnamefont {Xing}},
  \bibinfo {author} {\bibfnamefont {R.}~\bibnamefont {Nepal}}, \bibinfo
  {author} {\bibfnamefont {D.~L.}\ \bibnamefont {Gong}}, \bibinfo {author}
  {\bibfnamefont {X.}~\bibnamefont {Gui}}, \bibinfo {author} {\bibfnamefont
  {W.~W.}\ \bibnamefont {Xie}}, \bibinfo {author} {\bibfnamefont {D.~P.}\
  \bibnamefont {Young}}, \bibinfo {author} {\bibfnamefont {E.~W.}\ \bibnamefont
  {Plummer}},\ and\ \bibinfo {author} {\bibfnamefont {R.~Y.}\ \bibnamefont
  {Jin}},\ }\bibfield  {title} {\bibinfo {title} {Evidence for topological
  semimetallicity in a chain-compound {TaSe$_3$}},\ }\href
  {https://doi.org/https://doi.org/10.1038/s41535-020-00257-7} {\bibfield
  {journal} {\bibinfo  {journal} {npj Quantum Mater.}\ }\textbf {\bibinfo
  {volume} {5}},\ \bibinfo {pages} {53} (\bibinfo {year} {2020})}\BibitemShut
  {NoStop}%
\bibitem [{\citenamefont {Xu}\ \emph {et~al.}(2021)\citenamefont {Xu},
  \citenamefont {Han}, \citenamefont {Wang}, \citenamefont {Thoutam},
  \citenamefont {Pate}, \citenamefont {Li}, \citenamefont {Zhang},
  \citenamefont {Wang}, \citenamefont {Fotovat}, \citenamefont {Welp},
  \citenamefont {Zhou}, \citenamefont {Kwok}, \citenamefont {Chung},
  \citenamefont {Kanatzidis},\ and\ \citenamefont {Xiao}}]{extendedKohler}%
  \BibitemOpen
  \bibfield  {author} {\bibinfo {author} {\bibfnamefont {J.}~\bibnamefont
  {Xu}}, \bibinfo {author} {\bibfnamefont {F.}~\bibnamefont {Han}}, \bibinfo
  {author} {\bibfnamefont {T.}~\bibnamefont {Wang}}, \bibinfo {author}
  {\bibfnamefont {L.~R.}\ \bibnamefont {Thoutam}}, \bibinfo {author}
  {\bibfnamefont {S.~E.}\ \bibnamefont {Pate}}, \bibinfo {author}
  {\bibfnamefont {M.}~\bibnamefont {Li}}, \bibinfo {author} {\bibfnamefont
  {X.}~\bibnamefont {Zhang}}, \bibinfo {author} {\bibfnamefont
  {Y.}~\bibnamefont {Wang}}, \bibinfo {author} {\bibfnamefont {R.}~\bibnamefont
  {Fotovat}}, \bibinfo {author} {\bibfnamefont {U.}~\bibnamefont {Welp}},
  \bibinfo {author} {\bibfnamefont {X.}~\bibnamefont {Zhou}}, \bibinfo {author}
  {\bibfnamefont {W.~K.}\ \bibnamefont {Kwok}}, \bibinfo {author}
  {\bibfnamefont {D.~Y.}\ \bibnamefont {Chung}}, \bibinfo {author}
  {\bibfnamefont {M.~G.}\ \bibnamefont {Kanatzidis}},\ and\ \bibinfo {author}
  {\bibfnamefont {Z.-L.}\ \bibnamefont {Xiao}},\ }\bibfield  {title} {\bibinfo
  {title} {Extended kohler's rule of magnetoresistance},\ }\href
  {https://doi.org/10.1103/PhysRevX.11.041029} {\bibfield  {journal} {\bibinfo
  {journal} {Phys. Rev. X}\ }\textbf {\bibinfo {volume} {11}},\ \bibinfo
  {pages} {041029} (\bibinfo {year} {2021})}\BibitemShut {NoStop}%
\bibitem [{\citenamefont {Shoenberg}(1984)}]{LK}%
  \BibitemOpen
  \bibfield  {author} {\bibinfo {author} {\bibfnamefont {D.}~\bibnamefont
  {Shoenberg}},\ }\bibfield  {title} {\bibinfo {title} {Magnetic oscillations
  in metals},\ }\href
  {https://doi.org/https://doi.org/10.1017/CBO9780511897870} {\bibfield
  {journal} {\bibinfo  {journal} {Cambridge Univ. Press, Cambridge, England}\
  }\textbf {\bibinfo {volume} {97}},\ \bibinfo {pages} {235132} (\bibinfo
  {year} {1984})}\BibitemShut {NoStop}%
\bibitem [{\citenamefont {Mikitik}\ and\ \citenamefont
  {Sharlai}(1999)}]{Berry}%
  \BibitemOpen
  \bibfield  {author} {\bibinfo {author} {\bibfnamefont {G.~P.}\ \bibnamefont
  {Mikitik}}\ and\ \bibinfo {author} {\bibfnamefont {Y.~V.}\ \bibnamefont
  {Sharlai}},\ }\bibfield  {title} {\bibinfo {title} {Manifestation of berry's
  phase in metal physics},\ }\href
  {https://doi.org/https://doi.org/10.1103/PhysRevLett.82.2147} {\bibfield
  {journal} {\bibinfo  {journal} {Phys. Rev. Lett.}\ }\textbf {\bibinfo
  {volume} {82}},\ \bibinfo {pages} {2147} (\bibinfo {year}
  {1999})}\BibitemShut {NoStop}%
\bibitem [{\citenamefont {Taskin}\ and\ \citenamefont
  {Ando}(2011)}]{Berry_deviation}%
  \BibitemOpen
  \bibfield  {author} {\bibinfo {author} {\bibfnamefont {A.~A.}\ \bibnamefont
  {Taskin}}\ and\ \bibinfo {author} {\bibfnamefont {Y.}~\bibnamefont {Ando}},\
  }\bibfield  {title} {\bibinfo {title} {Berry phase of nonideal dirac fermions
  in topological insulators},\ }\href
  {https://doi.org/10.1103/PhysRevB.84.035301} {\bibfield  {journal} {\bibinfo
  {journal} {Phys. Rev. B}\ }\textbf {\bibinfo {volume} {84}},\ \bibinfo
  {pages} {035301} (\bibinfo {year} {2011})}\BibitemShut {NoStop}%
\bibitem [{\citenamefont {Hu}\ \emph {et~al.}(2017)\citenamefont {Hu},
  \citenamefont {Tang}, \citenamefont {Liu}, \citenamefont {Zhu}, \citenamefont
  {Wei},\ and\ \citenamefont {Mao}}]{LL}%
  \BibitemOpen
  \bibfield  {author} {\bibinfo {author} {\bibfnamefont {J.}~\bibnamefont
  {Hu}}, \bibinfo {author} {\bibfnamefont {Z.~J.}\ \bibnamefont {Tang}},
  \bibinfo {author} {\bibfnamefont {J.~Y.}\ \bibnamefont {Liu}}, \bibinfo
  {author} {\bibfnamefont {Y.~L.}\ \bibnamefont {Zhu}}, \bibinfo {author}
  {\bibfnamefont {J.}~\bibnamefont {Wei}},\ and\ \bibinfo {author}
  {\bibfnamefont {Z.~Q.}\ \bibnamefont {Mao}},\ }\bibfield  {title} {\bibinfo
  {title} {Nearly massless \textsc{D}irac fermions and strong \textsc{Z}eeman
  splitting in the nodal-line semimetal {ZrSiS} probed by de \textsc{H}aas-van
  \textsc{A}lphen quantum oscillations},\ }\href
  {https://doi.org/https://doi.org/10.1103/PhysRevB.96.045127} {\bibfield
  {journal} {\bibinfo  {journal} {Phys. Rev. B}\ }\textbf {\bibinfo {volume}
  {96}},\ \bibinfo {pages} {045127} (\bibinfo {year} {2017})}\BibitemShut
  {NoStop}%
\bibitem [{\citenamefont {Hu}\ \emph {et~al.}(2016)\citenamefont {Hu},
  \citenamefont {Tang}, \citenamefont {Liu}, \citenamefont {Liu}, \citenamefont
  {Zhu}, \citenamefont {Graf}, \citenamefont {Myhro}, \citenamefont {Tran},
  \citenamefont {Lau}, \citenamefont {Wei},\ and\ \citenamefont
  {Mao}}]{Mao-prl}%
  \BibitemOpen
  \bibfield  {author} {\bibinfo {author} {\bibfnamefont {J.}~\bibnamefont
  {Hu}}, \bibinfo {author} {\bibfnamefont {Z.~J.}\ \bibnamefont {Tang}},
  \bibinfo {author} {\bibfnamefont {J.~Y.}\ \bibnamefont {Liu}}, \bibinfo
  {author} {\bibfnamefont {X.}~\bibnamefont {Liu}}, \bibinfo {author}
  {\bibfnamefont {Y.~L.}\ \bibnamefont {Zhu}}, \bibinfo {author} {\bibfnamefont
  {D.}~\bibnamefont {Graf}}, \bibinfo {author} {\bibfnamefont {K.}~\bibnamefont
  {Myhro}}, \bibinfo {author} {\bibfnamefont {S.}~\bibnamefont {Tran}},
  \bibinfo {author} {\bibfnamefont {C.~N.}\ \bibnamefont {Lau}}, \bibinfo
  {author} {\bibfnamefont {J.}~\bibnamefont {Wei}},\ and\ \bibinfo {author}
  {\bibfnamefont {Z.~Q.}\ \bibnamefont {Mao}},\ }\bibfield  {title} {\bibinfo
  {title} {Evidence of topological nodal-line fermions in {ZrSiSe} and
  {ZrSiTe}},\ }\href
  {https://doi.org/https://doi.org/10.1103/PhysRevLett.117.016602} {\bibfield
  {journal} {\bibinfo  {journal} {Phys. Rev. Lett.}\ }\textbf {\bibinfo
  {volume} {117}},\ \bibinfo {pages} {016602} (\bibinfo {year}
  {2016})}\BibitemShut {NoStop}%
\bibitem [{\citenamefont {Fu}\ and\ \citenamefont
  {Kane}(2007)}]{PhysRevB.76.045302}%
  \BibitemOpen
  \bibfield  {author} {\bibinfo {author} {\bibfnamefont {L.}~\bibnamefont
  {Fu}}\ and\ \bibinfo {author} {\bibfnamefont {C.~L.}\ \bibnamefont {Kane}},\
  }\bibfield  {title} {\bibinfo {title} {Topological insulators with inversion
  symmetry},\ }\href {https://doi.org/10.1103/PhysRevB.76.045302} {\bibfield
  {journal} {\bibinfo  {journal} {Phys. Rev. B}\ }\textbf {\bibinfo {volume}
  {76}},\ \bibinfo {pages} {045302} (\bibinfo {year} {2007})}\BibitemShut
  {NoStop}%
\bibitem [{\citenamefont {Mostofi}\ \emph {et~al.}(2008)\citenamefont
  {Mostofi}, \citenamefont {Yates}, \citenamefont {Lee}, \citenamefont {Souza},
  \citenamefont {Vanderbilt},\ and\ \citenamefont {Marzari}}]{MOSTOFI2008685}%
  \BibitemOpen
  \bibfield  {author} {\bibinfo {author} {\bibfnamefont {A.~A.}\ \bibnamefont
  {Mostofi}}, \bibinfo {author} {\bibfnamefont {J.~R.}\ \bibnamefont {Yates}},
  \bibinfo {author} {\bibfnamefont {Y.-S.}\ \bibnamefont {Lee}}, \bibinfo
  {author} {\bibfnamefont {I.}~\bibnamefont {Souza}}, \bibinfo {author}
  {\bibfnamefont {D.}~\bibnamefont {Vanderbilt}},\ and\ \bibinfo {author}
  {\bibfnamefont {N.}~\bibnamefont {Marzari}},\ }\bibfield  {title} {\bibinfo
  {title} {wannier90: A tool for obtaining maximally-localised wannier
  functions},\ }\href
  {https://doi.org/https://doi.org/10.1016/j.cpc.2007.11.016} {\bibfield
  {journal} {\bibinfo  {journal} {Comput. Phys Commun.}\ }\textbf {\bibinfo
  {volume} {178}},\ \bibinfo {pages} {685} (\bibinfo {year}
  {2008})}\BibitemShut {NoStop}%
\bibitem [{\citenamefont {Sancho}\ \emph {et~al.}(1985)\citenamefont {Sancho},
  \citenamefont {Sancho}, \citenamefont {Sancho},\ and\ \citenamefont
  {Rubio}}]{Sancho_1985}%
  \BibitemOpen
  \bibfield  {author} {\bibinfo {author} {\bibfnamefont {M.~L.}\ \bibnamefont
  {Sancho}}, \bibinfo {author} {\bibfnamefont {J.~L.}\ \bibnamefont {Sancho}},
  \bibinfo {author} {\bibfnamefont {J.~L.}\ \bibnamefont {Sancho}},\ and\
  \bibinfo {author} {\bibfnamefont {J.}~\bibnamefont {Rubio}},\ }\bibfield
  {title} {\bibinfo {title} {Highly convergent schemes for the calculation of
  bulk and surface green functions},\ }\href
  {https://doi.org/10.1088/0305-4608/15/4/009} {\bibfield  {journal} {\bibinfo
  {journal} {J. Phys. F}\ }\textbf {\bibinfo {volume} {15}},\ \bibinfo {pages}
  {851} (\bibinfo {year} {1985})}\BibitemShut {NoStop}%
\bibitem [{\citenamefont {Wu}\ \emph {et~al.}(2018{\natexlab{b}})\citenamefont
  {Wu}, \citenamefont {Zhang}, \citenamefont {Song}, \citenamefont {Troyer},\
  and\ \citenamefont {Soluyanov}}]{WU2018405}%
  \BibitemOpen
  \bibfield  {author} {\bibinfo {author} {\bibfnamefont {Q.}~\bibnamefont
  {Wu}}, \bibinfo {author} {\bibfnamefont {S.}~\bibnamefont {Zhang}}, \bibinfo
  {author} {\bibfnamefont {H.-F.}\ \bibnamefont {Song}}, \bibinfo {author}
  {\bibfnamefont {M.}~\bibnamefont {Troyer}},\ and\ \bibinfo {author}
  {\bibfnamefont {A.~A.}\ \bibnamefont {Soluyanov}},\ }\bibfield  {title}
  {\bibinfo {title} {Wanniertools: An open-source software package for novel
  topological materials},\ }\href
  {https://doi.org/https://doi.org/10.1016/j.cpc.2017.09.033} {\bibfield
  {journal} {\bibinfo  {journal} {Comput. Phys. Commun.}\ }\textbf {\bibinfo
  {volume} {224}},\ \bibinfo {pages} {405} (\bibinfo {year}
  {2018}{\natexlab{b}})}\BibitemShut {NoStop}%
\end{thebibliography}%

\end{document}